\documentclass[12pt]{article}
\usepackage{publication}
\usepackage{graphicx}
\usepackage{upgreek}
\usepackage{multicol,multirow}
\usepackage{amsmath,amssymb,amsfonts}
\usepackage{mathrsfs}
\usepackage{amsthm}
\usepackage[figuresright]{rotating}
\usepackage{appendix}
\usepackage[authoryear]{natbib}
\usepackage{ifpdf}
\usepackage[T1]{fontenc}
\usepackage{newtxtext}
\usepackage{newtxmath}
\usepackage{textcomp}
\usepackage{xcolor}
\usepackage{subcaption}
\usepackage{siunitx}
\usepackage{placeins}

\title{The Role of Dynamic Stall in Aerofoil Shape Optimisation for Curvilinear Blade Kinematics}

\author{Benjamin Irwin$^{1*}$}
\contact{$^*$B.Irwin@soton.ac.uk}

\author{David Toal$^1$}

\author{Swathi Krishna$^1$}

\affil{$^1$Department of Aeronautics and Astronautics, University of Southampton, Southampton, SO17 1BJ, UK}

\date{\today}

\begin{document}

	\maketitle
	
	\begin{abstract}
		\footnotesize
		This study investigates the influence of aerofoil shape optimisation on blade aerodynamic performance under curvilinear and unsteady kinematics characteristic of vertical-axis turbines and cycloidal propellers. Using a cyclorotor in hover as a representative configuration, aerofoil optimisation was performed using two-dimensional unsteady Reynolds-averaged Navier–Stokes simulations coupled with Kriging. The optimised design was subsequently experimentally assessed through force measurements and flow-field characterisation using particle image velocimetry. Performance was enhanced through the suppression of leading-edge vortex separation during the primary thrust peak. This finding also reveals a governing constraint: the effectiveness of aerofoil optimisation depends on dynamic stall severity. Under light dynamic stall,  geometric modification promotes vortex attachment and improves aerodynamic loading. Under deep dynamic stall, flow separation dominates the blade aerodynamics, and aerofoil shape modification cannot suppress leading-edge vortex shedding.  The stall severity is regulated by rotor solidity through its influence on the induced throughflow-to-blade-speed ratio and the resulting effective incidence. Aerofoil optimisation is therefore viable primarily in high-solidity configurations that operate within a moderated stall regime. These findings establish a physics-based condition for aerofoil optimisation in curvilinear dynamic stall environments.

	\end{abstract}

	\section{Introduction}
	
	Recent years have seen the emergence of vertical axis turbines (VATs) and cycloidal propellers (cyclorotors), devices that share common aerodynamic principles yet serve distinct applications in renewable energy and propulsion. Both configurations comprise a rotating cylindrical arrangement of blades. VATs can offer improved performance in high-density wind farms and operate effectively in highly variable wind conditions, despite often being less efficient than conventional horizontal-axis designs in standalone operation \citep{Hansen2021NumericalApproach}. Cyclorotors are employed widely in Voith-Schneider propellers mounted on large ships \citep{Voith2023VoithVSP}. These propellers leverage the thrust-vectoring capabilities of cyclorotors to enhance manoeuvrability, particularly during docking. While the use of cyclorotors in the maritime industry is not new, renewed interest  for micro-air vehicles (MAVs) applications has emerged. At typical Reynolds numbers of 10,000-40,000, cyclorotors have demonstrated lower susceptibility than conventional rotors to the detrimental low-speed effects \citep{Shrestha2017DevelopmentApplication}.

\begin{figure}[h]
		\centering
		\includegraphics{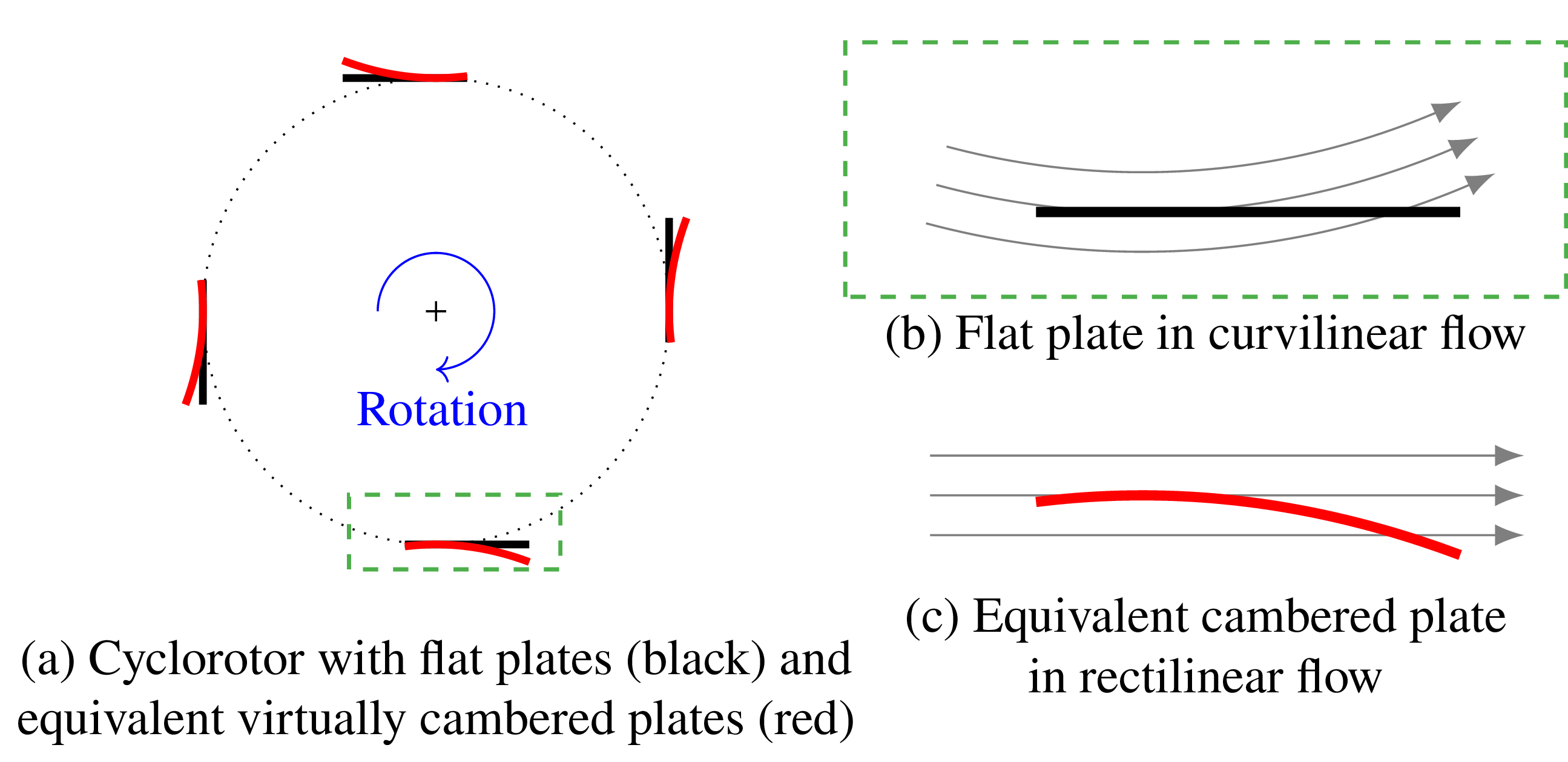}
		\caption{Virtual camber effect on a cyclorotor} \label{fig:virtual_camber_cyclo}
		\phantomsubcaption\label{fig:virtual_camber_cyclo_A}
		\phantomsubcaption\label{fig:virtual_camber_cyclo_B}
		\phantomsubcaption\label{fig:virtual_camber_cyclo_C}
	\end{figure}

	The rotating, cylindrical arrangement subjects the blades to a curvilinear flowfield characterised by curved streamlines and introduces associated aerodynamic phenomena. A principle feature of the flowfield around a blade in curvilinear flow is the chordwise variation in local-angle-of-attack. This variation results in a phenomenon known as virtual camber, wherein the blade's aerodynamic behaviour can be approximated by that of a cambered aerofoil in rectilinear flow that maintains the same localised chordwise incidence \citep{Migliore1980FlowAerodynamics} (Fig. \ref{fig:virtual_camber_cyclo}). For this effectively cambered shape, the zero-lift angle-of-attack ($\alpha_{0}$) can be determined and subsequently used to offset the geometric blade pitch ($\alpha_{geom}$), yielding an effective angle-of-attack ($\alpha_{eff}$) (Eq. \ref{eq:virtual_camber}).

	\begin{equation}
		\alpha_{eff} = \alpha_{geom} - (\alpha_0)_{\text{virtual camber}}
		\label{eq:virtual_camber}
	\end{equation}
	
	In addition to curvilinear effects, the rotating cylindrical arrangement of blades imposes highly unsteady flow conditions. This unsteadiness frequently induces the aerodynamic phenomena known as dynamic stall, wherein unsteady blade kinematics enable transient lift generation well beyond its static stall angle \citep{Mulleners2012TheRevisited}. Dynamic stall is often accompanied by the formation of leading-edge vortices (LEVs), which arise as the boundary layer separates from the blade's leading edge and rolls up into a large vortex close to the blade surface. Consequently, the resulting flowfields are heavily vortex-dominated.

\begin{figure}[h]
		\centering
		\includegraphics{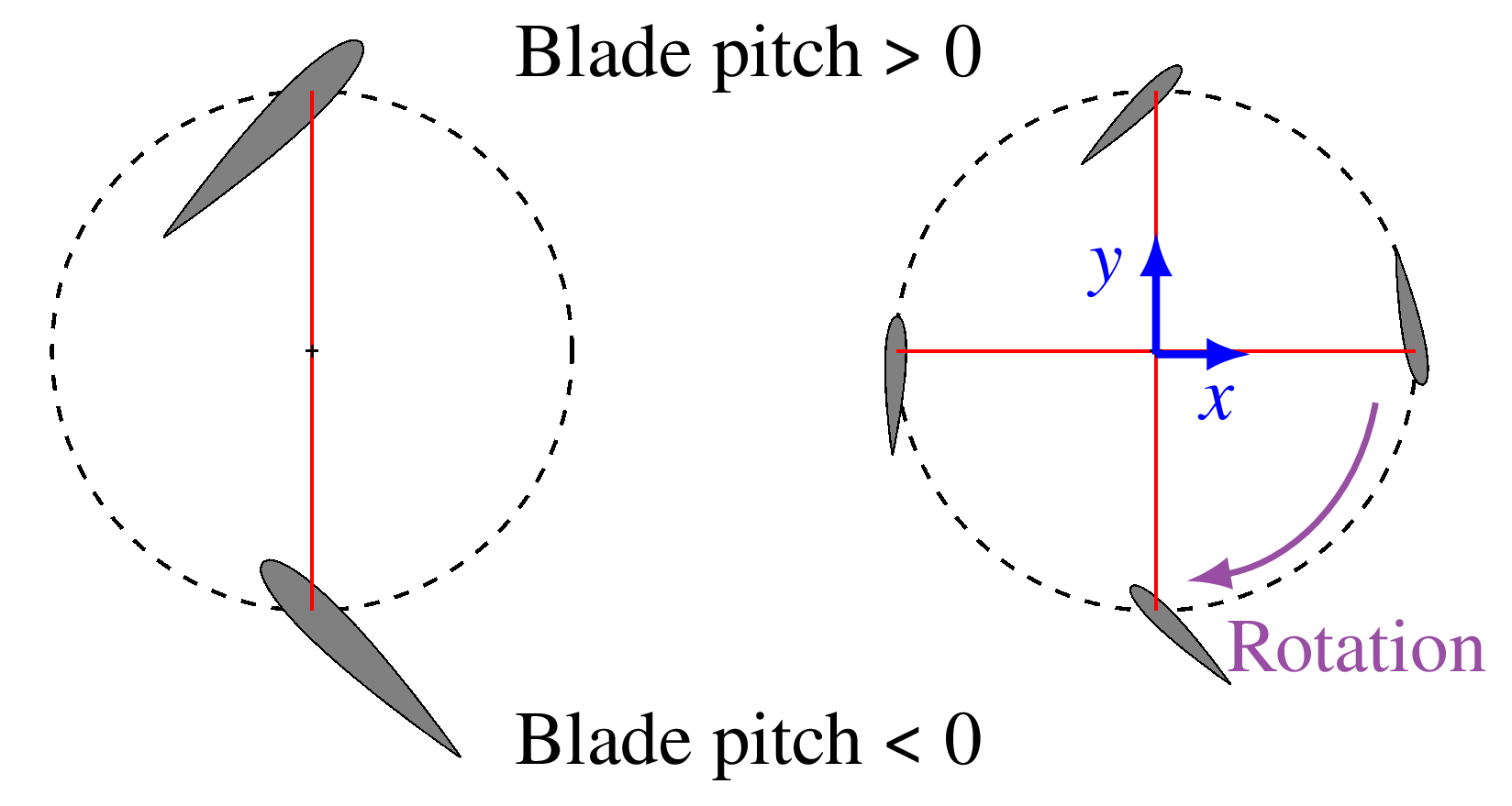}
		\caption{Two rotors of fixed solidity ($\sigma=0.3395$) with (left) two blades with chord $c$ and (right) four blades with chord $c/2$ }
		\label{fig:solidity_example}
	\end{figure}

	Cyclorotors present a compelling case for aerofoil optimisation, due to an inherent  design trade-off linked to rotor solidity ($\sigma$), which represents the proportion of the rotor's circumference that is occupied by blades. Rotor solidity is calculated using blade-count ($N_b$), blade chord ($c$) and rotor radius ($R$) (Eq. \ref{eq:solidity}). For a fixed blade pitch amplitude and RPM,  there will generally be some minimum level of solidity required to reach a target level of thrust. The means by which this solidity is realised introduces competing aerodynamic consequences (Fig. \ref{fig:solidity_example}). Configurations employing high blade counts with shorter chord lengths produce a consistent thrust output, but typically suffer from reduced efficiency. 
	\begin{equation}
		\sigma = \frac{N_b c}{2 \pi R}
		\label{eq:solidity}
	\end{equation}
	
	This reduction in efficiency has been observed in both cyclorotors and VATs and is generally attributed to blade-to-blade interaction effects, arising from reduced inter-blade spacing, which promotes wake interaction between successive blades \citep{Benedict2010FundamentalApplications,Zhang2025EffectTurbines}. Cyclorotors additionally will often operate at high blade-pitch amplitudes to exploit the increased lift coefficient possible under dynamic stall. This operating mode promotes frequent LEV shedding and blade-vortex interactions. Given the strong dependence between aerofoil shape and flow separation, there is significant potential for aerofoil optimisation to improve the performance of high blade-count cyclorotors, thereby fully leveraging their inherent steady thrust characteristics. The concept of using cambered aerofoils in cyclorotors and optimising their shape has been explored in previous studies and have shown that the performance of a chosen cyclorotor configuration could indeed be improved through an optimisation of the aerofoil shape \citep{Tang2017UnsteadyModel, Zhang2018ThePropellers}. The range of different aerofoil shapes used in these studies remain relatively restrictive and do not encompass geometries with leading-edge droops, which have been associated with additional performance benefits \citep{Ferrier2020InvestigationMorphing}.
	
	Overall, devices that utilise blades operating in curvilinear flow are subject to multiple complex and overlapping effects. Classical aerodynamic theory typically relies on the assumption of rectilinear 2D flows, which constrains the accurate interpretation and modelling of the blade aerodynamics in these environments. Such frameworks do not account for the coupled effects of virtual camber, dynamic stall, and blade–wake interaction inherent to curvilinear flows. Consequently, the aerofoil geometry-driven control of LEV behaviour, and the effectiveness of optimisation across different solidity remain unresolved, resulting in the absence of a clear set of physics-based design principles for achieving an optimum aerofoil shape.

	The aim of the current study is to investigate the governing physical mechanisms influencing aerofoil optimisation in a curvilinear flow. The study approaches this problem by optimising the aerofoil for a cyclorotor in the hover condition. The hover condition preserves the predominantly curvilinear inflow characteristics without the interference of the freestream effects, aligning with the primary focus of the investigation. Furthermore, in both maritime and aviation contexts, cyclorotors are frequently utilised in pure hover or at very low freestream speeds. To further examine the influence of rotor configuration on the resulting design principles,  aerofoil optimisation is conducted for cyclorotors with 1-4 blades. This process will have the additional benefit of potentially highlighting how high blade-count cyclorotors can be made more viable to exploit their inherently steadier thrust output.

	Although cyclorotors represent a specific manifestation of curvilinear blade motion, the optimisation outcomes are expected to retain relevance for  other devices such as VATs. Despite serving distinct functional roles, both configurations experience comparable governing aerodynamic phenomena, including virtual camber, wake intersection, and dynamic stall. Consequently, while the exact optimal geometries may differ between applications, the underlying design principles are expected to remain broadly consistent.

	\section{Methodology} \label{section:methods}
	\subsection{Cyclorotor configuration and parametrization} \label{section:basic_setup}
	
	A 4-bladed cyclorotor operating at a representative chord-based Reynolds number of 30,000 was chosen as the primary baseline configuration for aerofoil optimisation (see Table \ref{table:base_design} for details). In both simulation and experiment, the cyclorotor uses water as its working fluid.	In addition to the 4-blade configuration, blade-counts of 1-3 were also examined to assess the impact of blade-count on optimisation outcomes and to isolate the aerodynamic effects of aerofoil optimisation for both single-blade and multi-blade flow dynamics. A 4-bar linkage system is used to provide an approximately sinusoidal pitching profile between $\pm$45 \si{\degree} (See supplementary material) \citep{Cogan2022NumericalState}.

	\begin{table}[h]
		\centering
		\begin{tabular}{|l|l|}
			\hline
			Rotor Radius ($R$)   & 150 \si{\milli\metre}        \\ \hline
			Blade Chord ($c$)     & 80 \si{\milli\metre}          \\ \hline
			Blade Span ($b$)       & 200 \si{\milli\metre}         \\ \hline
			Number of Blades ($N_b$) & 1-4           \\ \hline
			Max/Min Pitch    & $\pm$45 \si{\degree}            \\ \hline
			RPM        & 24      \\ \hline
			Chord-based Reynolds Number ($Re=U_{b} c/\nu$)   & $\sim$30,000 \\ \hline
			Default Aerofoil	& NACA 0015 \\ \hline
		\end{tabular}
		\caption{Baseline cyclorotor design}
		\label{table:base_design}
	\end{table}

\begin{figure}[b]
		\centering
		\includegraphics{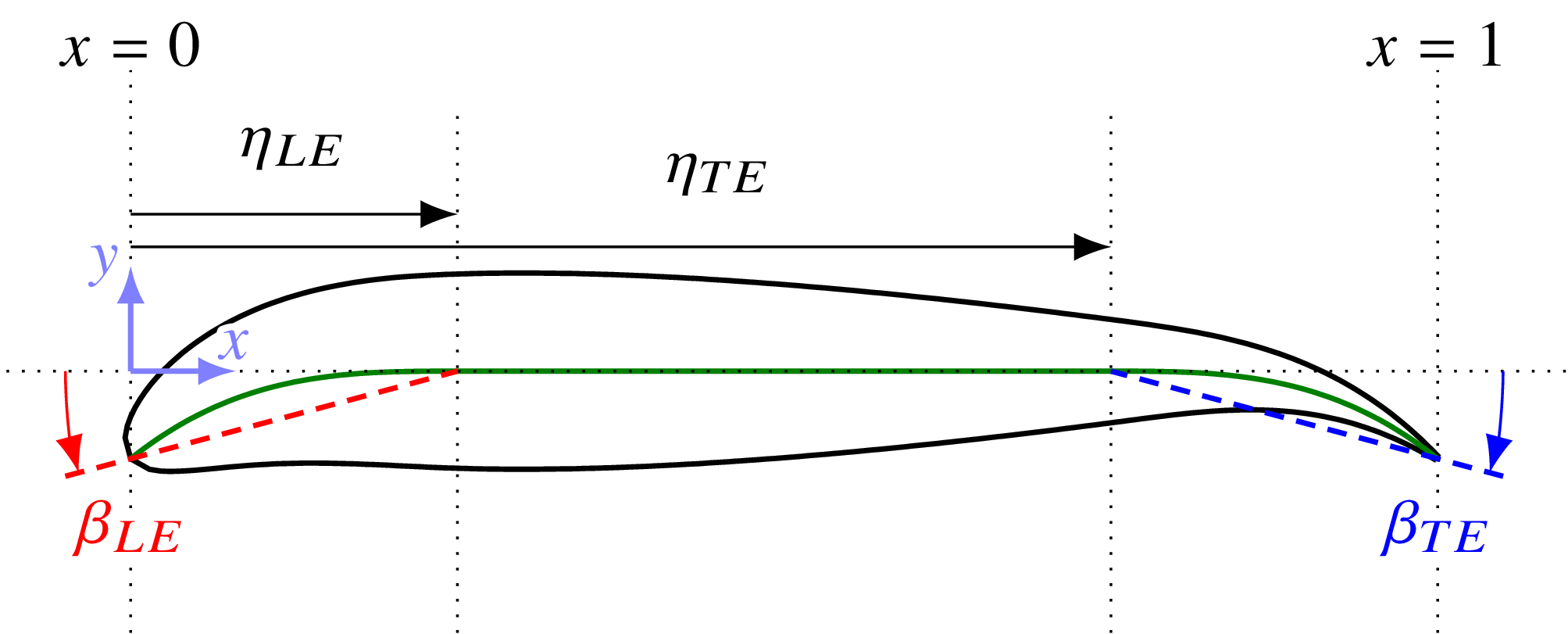}
		\caption{Blade-shape parametrization}
		\label{fig:parameters}
	\end{figure}

	For this baseline configuration, a NACA 0015 profile was selected as the reference aerofoil due to its widespread adoption in prior cyclorotor investigations, facilitating direct comparison of aerodynamic performance  \citep{Benedict2010FundamentalApplications, Reed2019ForceNumbers, Yu2016Two-dimensionalHover, Walther2019SymmetricNumbers, Jarugumilli2013AnFlight}. Additional aerofoil geometries were generated through camberline modification of the baseline profile via the introduction of a leading-edge (LE) and trailing-edge (TE) droop to the camberline. Each droop is defined by a chordwise hinge-point location ($\eta$) and droop angle ($\beta$), yielding four independent design variables governing the final aerofoil shape: $\eta_{\text{LE}}$, $\beta_{\text{LE}}$, $\eta_{\text{TE}}$ and $\beta_{\text{TE}}$ (Fig. \ref{fig:parameters}). The subsequent camberline is defined by Eq. \ref{eq:TE_camber}) \citep{Huang2021EvaluateRotor, Huang2022RelyingRotor, Woods2014AerodynamicConcept}. In contrast to the NACA series of aerofoils, this parametrization allows independent control of the leading- and trailing-edge geometry that is expected to strongly influence LEV behaviour, without introducing global camber changes that would complicate isolating their respective effects. Droop angles are limited to $\pm \ang{30}$, with hinge-point locations bounded by $\eta_{\text{LE}} \in [0.15c,0.5c]$ and $\eta_{\text{TE}} \in [0.5c,0.85c]$.

	The defined camberline was thickened with a 15\% thickness NACA profile to generate the complete aerofoil geometry (Eq. \ref{eq:NACA_thick}). Using the slope ($\theta$) and position along the camberline, the surface coordinates were obtained from Eqns.\ref{eq:NACA_thick}-\ref{eq:surface_coords} \citep{Abbott1959TheoryData}.

	\begin{equation}
		y_{c} =     
		\begin{cases}
			\frac{-\eta_{LE}\tan{\beta_{LE}}}{\eta_{LE}^{3}} (\eta_{LE} - x)^{3} & \text{if } 0 \leq x \leq \eta_{LE} \\[0.75em] 
			\frac{-(1-\eta_{TE})\tan{\beta_{TE}}}{(1-\eta_{TE})^{3}} (x - \eta_{TE})^{3} & \text{if } \eta_{TE} \leq x \leq 1 \\[0.75em] 
			0 & \text{otherwise}
		\end{cases}
		\label{eq:TE_camber}
	\end{equation}

	\begin{equation}
		y_t = 5 t (0.2969\sqrt{x} - 0.1260x - 0.3516x^2 + 0.2843x^3 - 0.1015x^4)
		\label{eq:NACA_thick}
	\end{equation}

	\begin{equation}
		[x, y]_{surface} = 
		\begin{cases}
			[x - y_t \sin(\theta), y_c + y_t \cos(\theta) ] & \text{upper surface} \\[1em]
			[x + y_t \sin(\theta), y_c - y_t \cos(\theta) ] & \text{lower surface} \\
		\end{cases}
		\label{eq:surface_coords}
	\end{equation}

	To characterise the unsteadiness of the aerodynamics in this setup, the non-dimensional reduced frequency ($k = \pi f c/U$) is used, which is expressed as a function of the blade chord $c$, flow velocity $U$, and the blade pitching frequency $f$. Flows with $0.05<k\leq0.2$ are generally classified as unsteady, while those with $k>0.2$ are considered highly unsteady \citep{Leishman2006PrinciplesAerodynamics}. For devices such as VATs and cyclorotors, the blade pitching frequency corresponds to the rotational frequency. The incident flow velocity experienced by the blades is primarily determined by the rotational speed of the cylindrical blade arrangement, ($U = 2 \pi f R$). As a result, the reduced frequency for these devices can often be expressed as $k = \frac{c}{2 R}$, a term solely dependent on the rotor's physical dimensions and independent of its pitching frequency or rotational speed. For the present configuration, the reduced frequency was found to be $k=0.27$ and thus would be classified as highly unsteady. While this exceeds values commonly observed in vertical-axis turbines ($k \leq 0.2$) \citep{Buchner2018DynamicConsiderations,RosadoHau2020ATurbines}, it accentuates the dynamic stall and vortex-dominated mechanisms central to this investigation.

	From this baseline configuration and aerofoil parametrization, new designs were generated for the optimisation process to identify the best performing aerofoil.

	\subsection{Optimisation via Kriging} \label{section:methods_kriging}

	The optimisation in this study was conducted using a Kriging surrogate model, a technique that has been used successfully in multiple other optimisation studies involving unsteady flows \citep{Tang2017UnsteadyModel,Choi2011EnhancementOptimization, Raul2021Surrogate-basedCriteria}. Unlike traditional regression or curve-fitting techniques, Kriging is a stochastic interpolation method that utilises a probabilistic framework to maximise the likelihood of the observed dataset. This foundation in probability allows for the uncertainty of the surrogate model to be estimated throughout the design space. This allows the use of the expected improvement ($EI$) update criterion, which attempts to balance design improvement with design exploration, consequently alleviating the risk of the design optimisation converging on a local minima/maxima rather than the global minimum/maximum \citep{Forrester2008EngineeringGuide}.

	An initial design space comprising of 20 aerofoil geometries was generated via a space-filling Latin hypercube. Subsequent update points were chosen with expected improvement criterion and the current global maximum of the surrogate model. Additional designs were iteratively added until no further improvement in the best evaluated design was observed after 6 consecutive updates. These methods were implemented in MATLAB using the toolbox from \cite{Forrester2008EngineeringGuide}.
	
	Aerofoil performance was quantified using the Figure of Merit ($FM$) metric, which assesses efficiency in hover. This metric can be expressed in terms of Disk Loading ($DL$) and Power Loading ($PL$) through momentum theory given in equations \ref{eq:pl}, \ref{eq:dl} and \ref{eq:fm} \citep{Benedict2015ExperimentalHover}. It represents the ratio of the ideal power required for hover to the actual power consumed. For interim analysis the coefficients of force and torque were evaluated, which are defined in terms of the blade speed ($U_b$) due to rotation and the fluid density ($\rho$) (Eq. \ref{eq:ft}). For the single-bladed results thrust and power coefficients were normalised by the blade planform area ($S = c \times b$), as opposed to the full rotor projected area (rotor diameter $\times$ blade span) typically employed in rotor-level performance assessment. For the four-bladed data, the full rotor projected area was used. The thrust direction is described with its angle ($\beta$) with respect to the vertical (Eq. \ref{eq:beta}).
	
	The Kriging surrogate framework was subsequently used to identify candidate aerofoil geometries for computational evaluation, with optimisation targeting maximisation of $FM$. The overall aim of the optimisation is identify a optimal aerofoil that could be tested experimentally to investigate the underpinning principles behind what makes an aerofoil optimal.

	\begin{equation}
		PL= \frac{\text{thrust}}{\text{power}}= \frac{\text{thrust}}{\text{torque}\cdot\text{RPM}\cdot\frac{2\pi}{60}},
		\label{eq:pl}
	\end{equation}
	\begin{equation}
		DL= \frac{\text{thrust}}{\text{rotor projected area}} = \frac{\text{thrust}}{2Rb},
		\label{eq:dl}
	\end{equation}
	\begin{equation}
		FM = PL\sqrt{\frac{DL}{2 \rho}},
		\label{eq:fm}
	\end{equation}
	\begin{equation}
		C_{T} = \frac{\text{thrust}}{\frac{1}{2}\rho U_b^2 S},\quad C_{torque} = \frac{\text{torque}}{\frac{1}{2}\rho U_b^2 S R}, \quad S = 
		\begin{cases}
			cb & N_b=1 \\
			2Rb & N_b>1
		\end{cases}, \quad U_b = RPM \cdot\frac{2\pi}{60} \cdot R,
		\label{eq:ft}
	\end{equation}
	\begin{equation}
		\beta = \arctan{\left(\frac{F_x}{F_y} \right)},
		\label{eq:beta}
	\end{equation}

	\subsection{Computational methodology} \label{section:methods_CFD}

	\begin{figure}[h]
		\centering
		
		\begin{subfigure}{0.49\linewidth}
			\centering
			\includegraphics{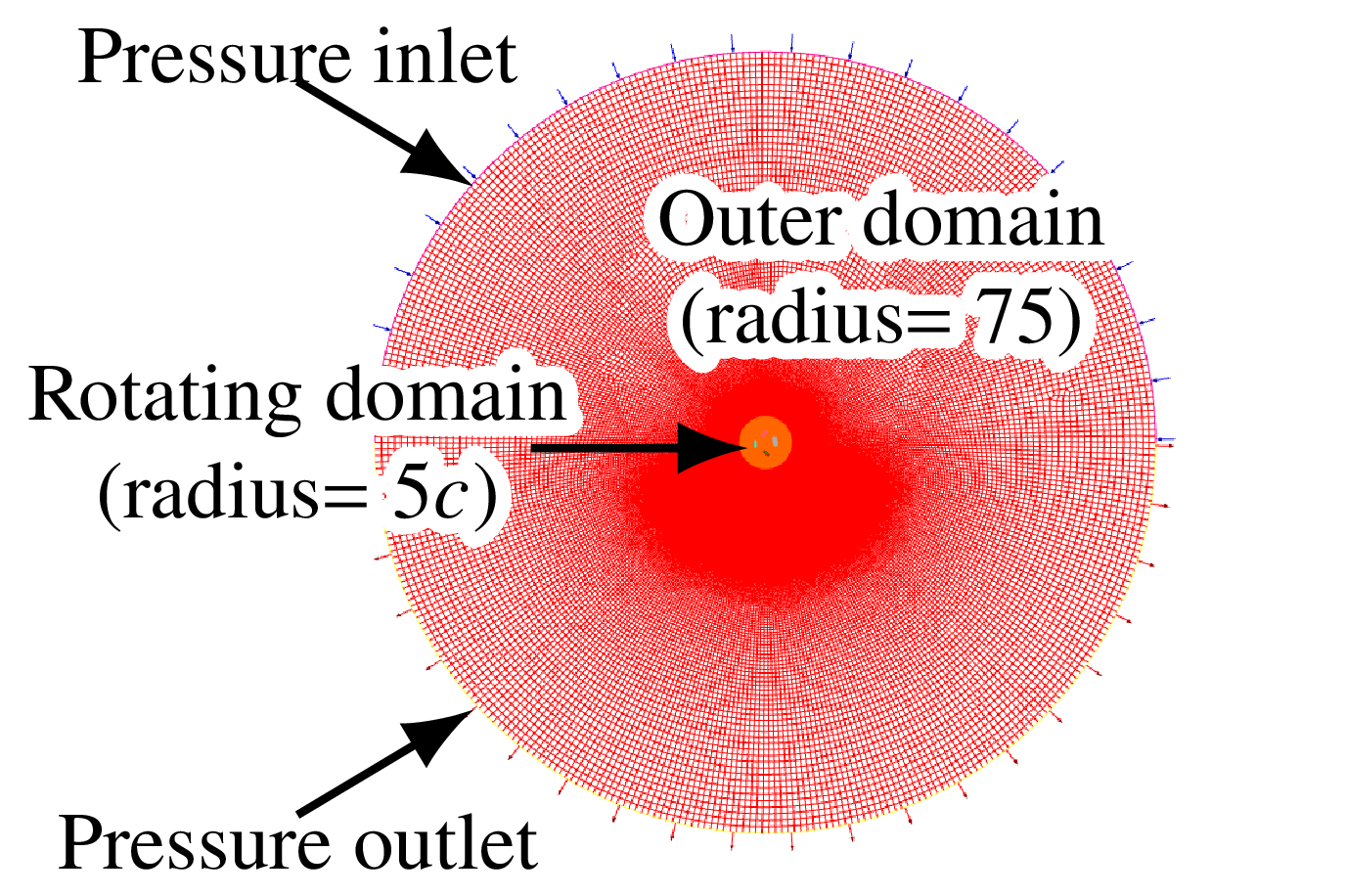}
			\label{fig:mesh_1}
		\end{subfigure}
		\begin{subfigure}{0.49\linewidth}
			\centering
			\includegraphics{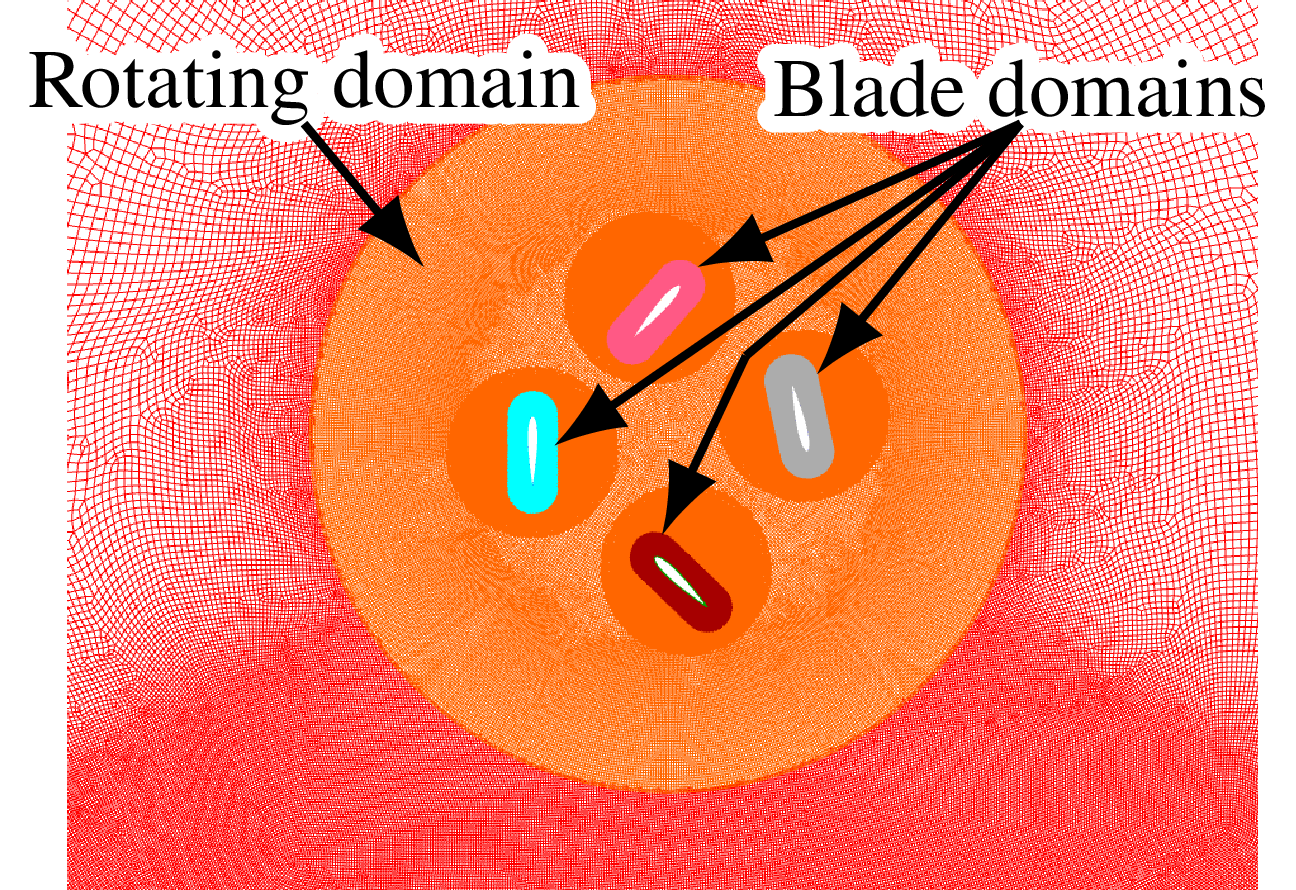}
			\label{fig:mesh_2}
		\end{subfigure}
		\begin{subfigure}{\linewidth}
			\centering
			\includegraphics{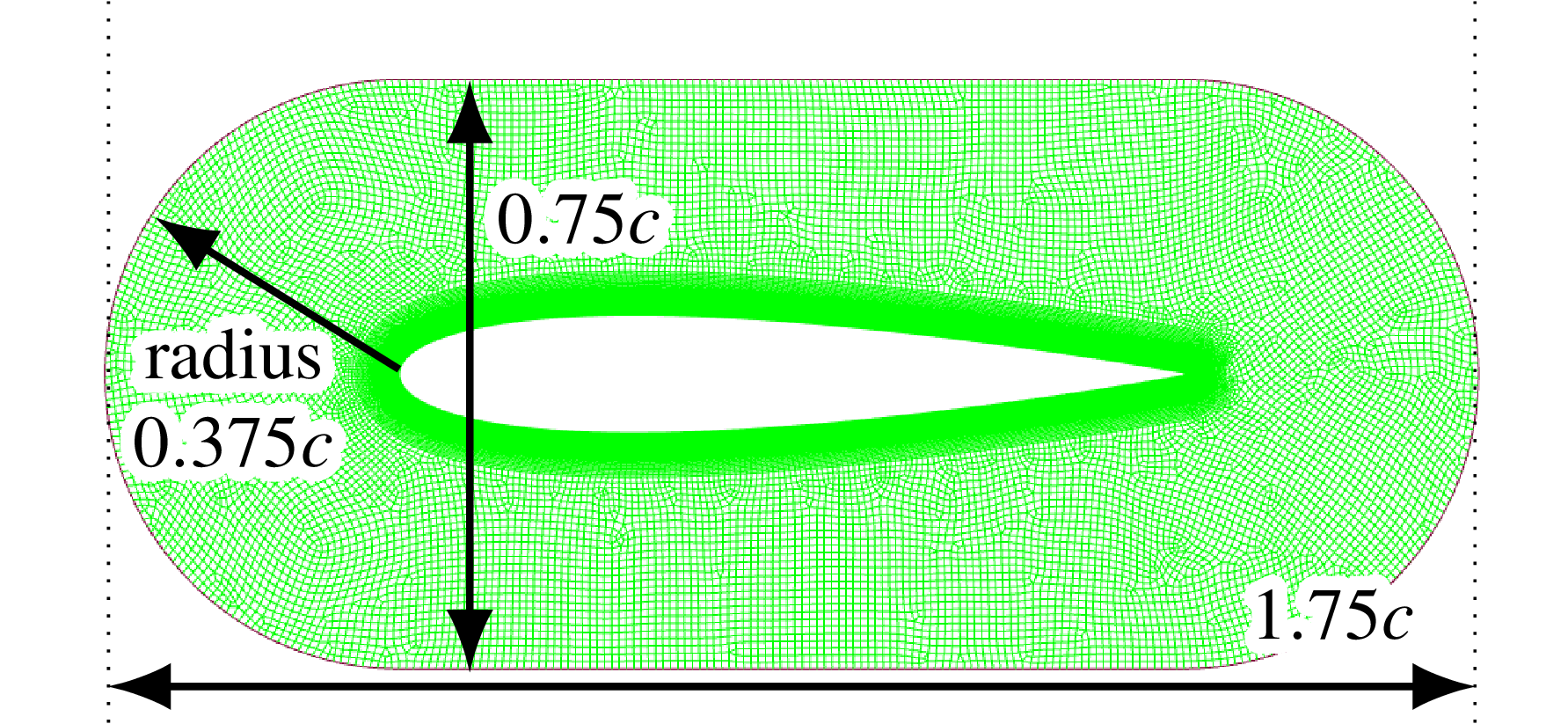}
		\end{subfigure}
		\caption{Overview of the mesh used for 2D CFD. ~800k cells after combination of outer, rotating and blade domains via overset meshing. Further details on mesh in Appendix \ref{appendix:URANS}.}
		\label{fig:CFD_mesh}
	\end{figure}

	To evaluate the performance of different aerofoil shapes within the chosen design space, computational fluid dynamics (CFD) was used, facilitating a preliminary assessment of design performance without requiring the fabrication of a physical model for each updated aerofoil geometry. Two-dimensional unsteady Reynolds-averaged Navier Stokes equations (URANS) were solved using the commercial software Ansys FLUENT. This approach has been successfully applied to cyclorotor simulations in previous studies \citep{Hu2016InvestigationSolver, Yun2007DesignCyclocopter, Xisto2017ParametricConditions, Zhang2018ThePropellers, Shi2022NumericalNumber, Yu2016Two-dimensionalHover, Hansen2021NumericalApproach, Ullah2022Two-DimensionalStall}. The limitations of URANS and chosen turbulence model are outlined in Appendix \ref{appendix:URANS}.

	The cyclorotor kinematics were simulated using separate meshes for the outer, rotating, and blade regions, that were then combined through overset meshing (Fig. \ref{fig:CFD_mesh}). The rotating domain was set to a constant angular velocity, while the blade domains were pitched periodically using a custom user-defined function (UDF) to match the experimental cyclorotor's kinematics. Each rotor revolution was resolved over 2000 timesteps. To ensure that the flow had reached a steadily periodic state, the simulations were run for 120 blade-passes, which was found to be sufficient for the 10-cycle averages to converge for cases using the baseline NACA 0015 aerofoil. The mean and phase-averaged forces/torques presented in this study were then taken over these last 10 revolutions.
	
	To accurately model boundary layer separation and dynamic stall, no wall functions were used, and the mesh was refined to ensure $y^+<1$. For the wider domain, the mesh was refined sufficiently to provide an average convective Courant number of $\approx0.5$. The $k-\omega$ SST turbulence model was selected for its proven suitability in flows experiencing dynamic stall, making it the most common choice for both vertical axis turbine (VAT) and cyclorotor simulations \citep{Hansen2021NumericalApproach, Hu2016InvestigationSolver, Zhang2018ThePropellers, Shi2022NumericalNumber, Yu2016Two-dimensionalHover, Ullah2022Two-DimensionalStall}.

	A single mesh was created for the blade-domain using the baseline NACA 0015. By deforming this baseline mesh, the meshes for subsequent aerofoil designs could be created.

	\subsubsection{Blade mesh deformation}
	
	\begin{figure}[h]
		\centering
		\includegraphics{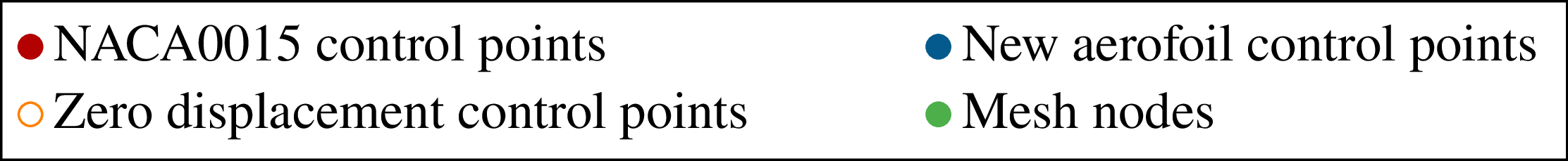}
		\begin{subfigure}{0.49\linewidth}
			\centering
			\includegraphics{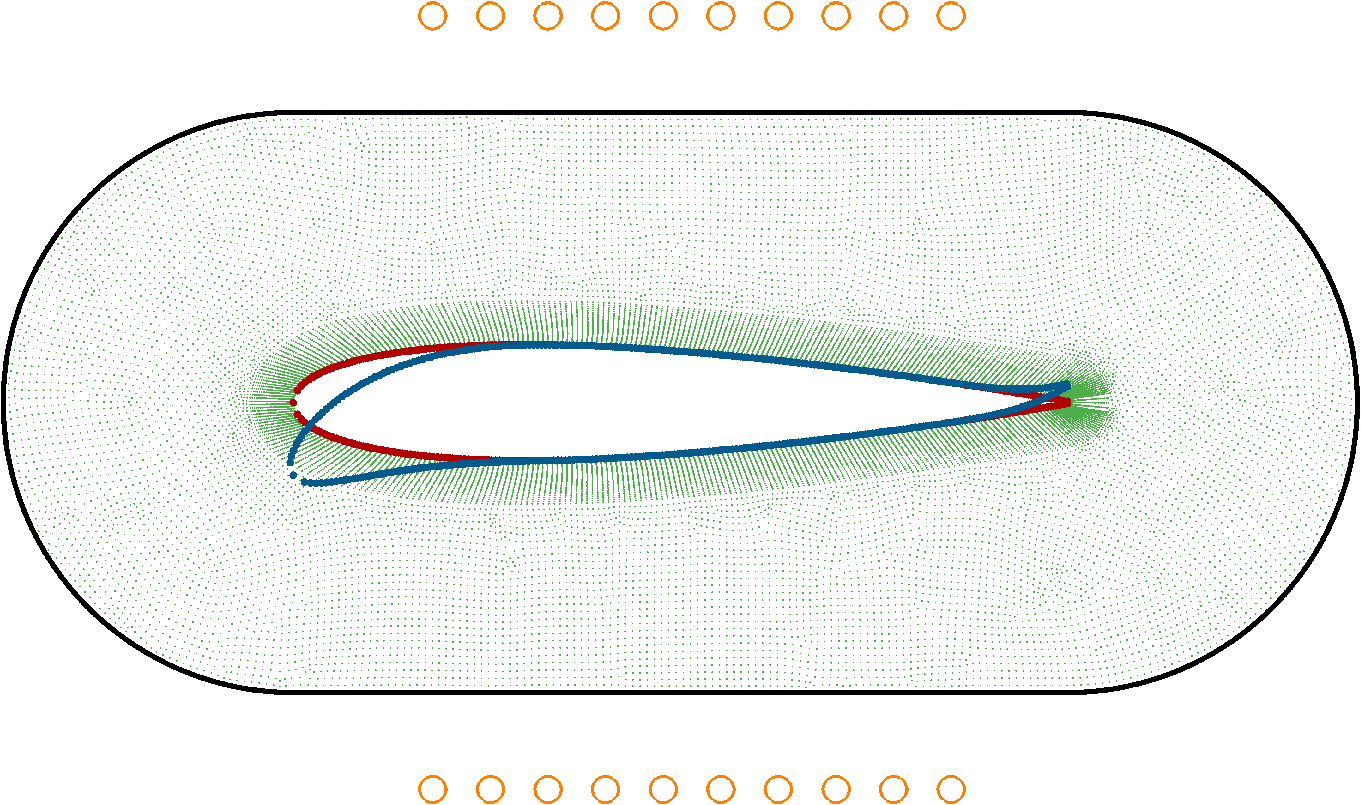}
			\caption{Setup of control points to define deformation map}
		\end{subfigure}
		\begin{subfigure}{0.49\linewidth}
			\centering
			\includegraphics{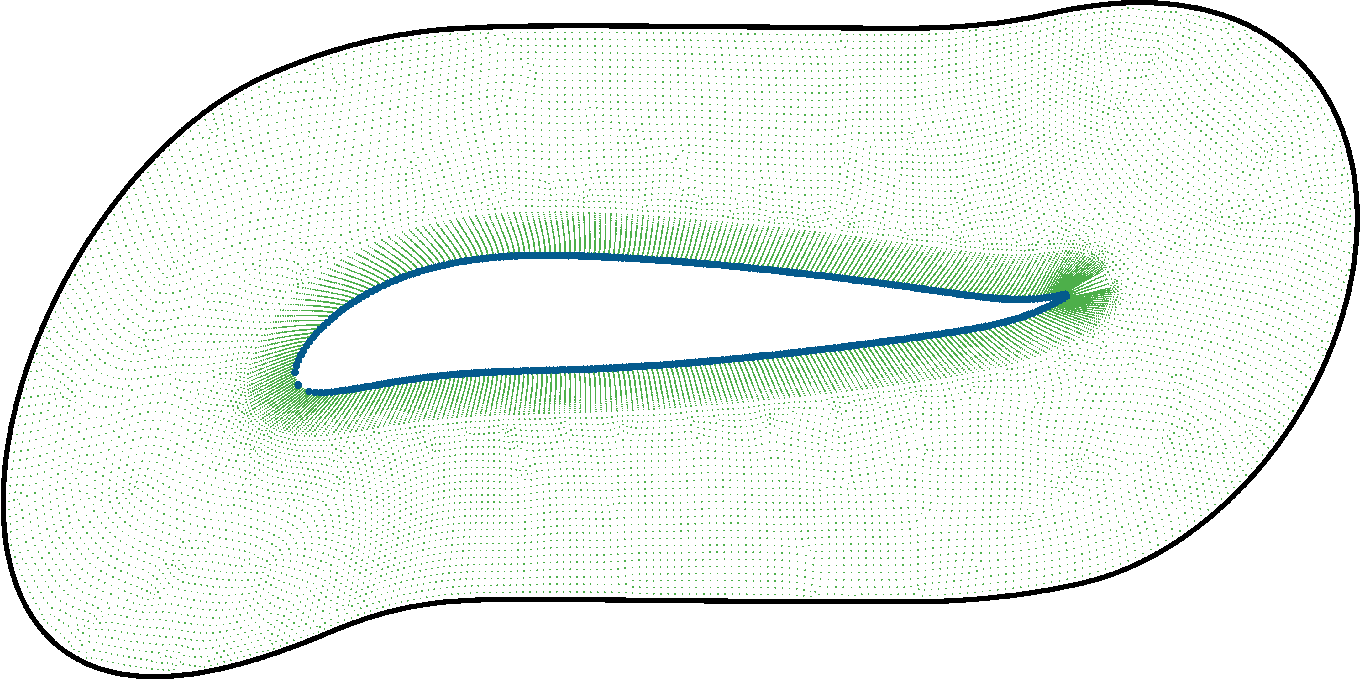}
			\caption{New mesh created by deforming mesh for NACA 0015}
		\end{subfigure}
		\caption{Demonstration of mesh deformation for new aerofoil shapes}
		\label{fig:mesh_deformation_demo}
	\end{figure}
	
	To simplify the creation of new meshes for different aerofoil shapes, the original mesh for the default NACA 0015 was deformed to create the meshes for new aerofoils (Fig. \ref{fig:mesh_deformation_demo}). This approach has been used in prior studies and keeps the mesh topology constant, consequently reducing the noise associated with varying mesh topology \citep{Toal2014OnQuantification, Huyse2002ProbabilisticUncertainty,Widhalm2007ComparisonDesign, Lassila2010ParametricMethod}. 
	
	For the original NACA 0015, a fixed set of positions on the camberline were used to generate a set of control points along the aerofoil's surface (Eq. \ref{eq:surface_coords}). A similar set of control points were then generated for the new aerofoil shape using the same set of positions on the camberline. This procedure yielded the required $x$ and $y$ displacement at each control point necessary to transform the baseline geometry into the updated aerofoil shape. Radial basis function interpolation was used to propagate the deformation in all areas of the surrounding domain \citep{Forrester2008EngineeringGuide}. Through this interpolation, the NACA 0015 mesh could then be deformed to produce a new mesh. To constrain these deformations to primarily regions near the aerofoil, 20 additional control points were placed in the farfield at which zero deformation was defined.

	\subsection{Experiments}
	
	\begin{figure}[h]
		\centering
		\includegraphics[width=\linewidth]{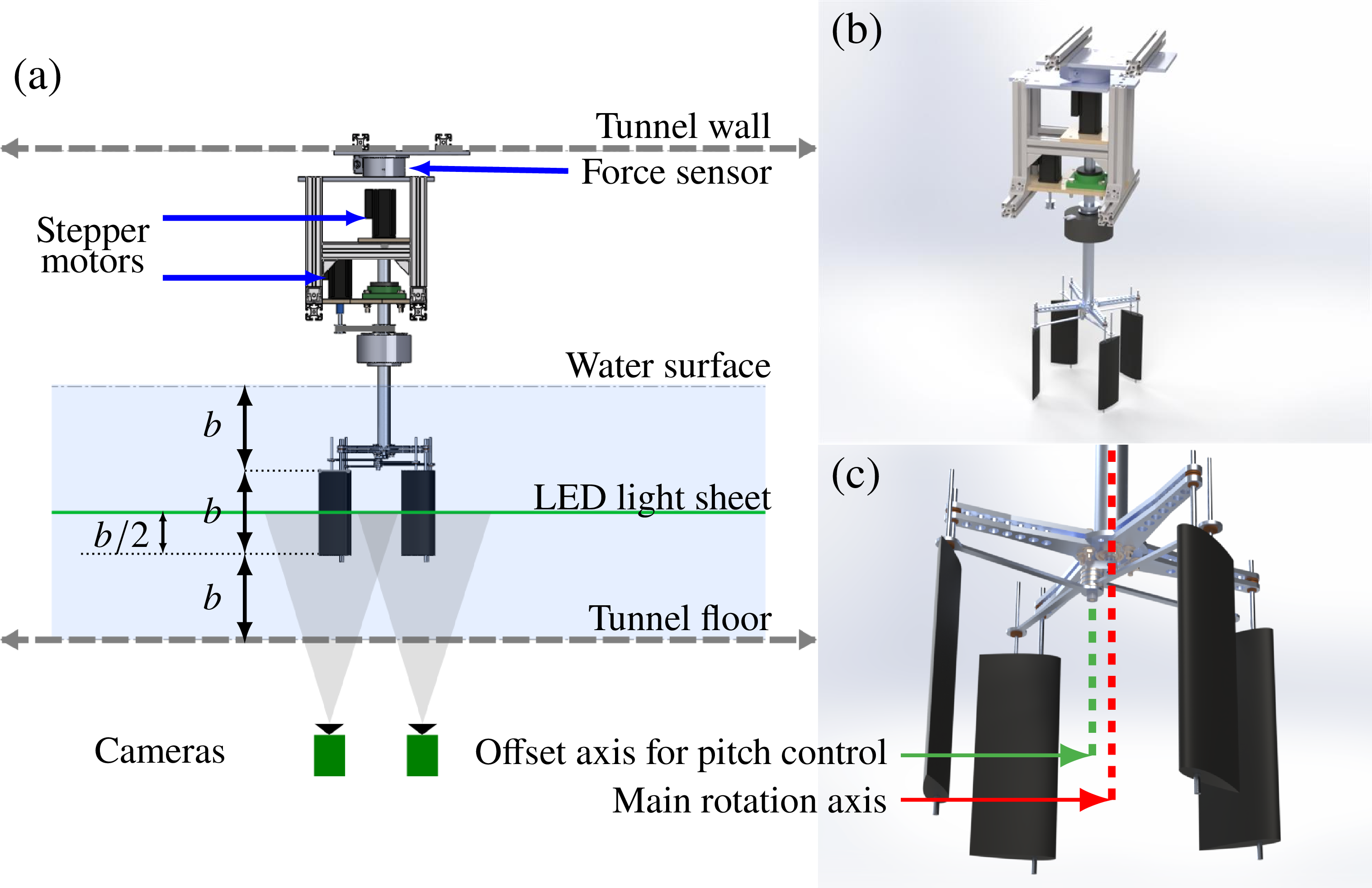}
		\caption{Setup of force and flowfield measurement of cyclorotor rig from (a) side view and (b) 3/4 view. (c) Close-up of 4-bar linkage system for blade pitch control}
		\label{fig:experimental_setup}
	\end{figure}
	\begin{figure}[h]
		\centering
		\begin{minipage}{0.49\linewidth}
			\begin{subfigure}{\linewidth}
				\centering
				\includegraphics[trim=0cm 1cm 0cm 1cm,clip=true]{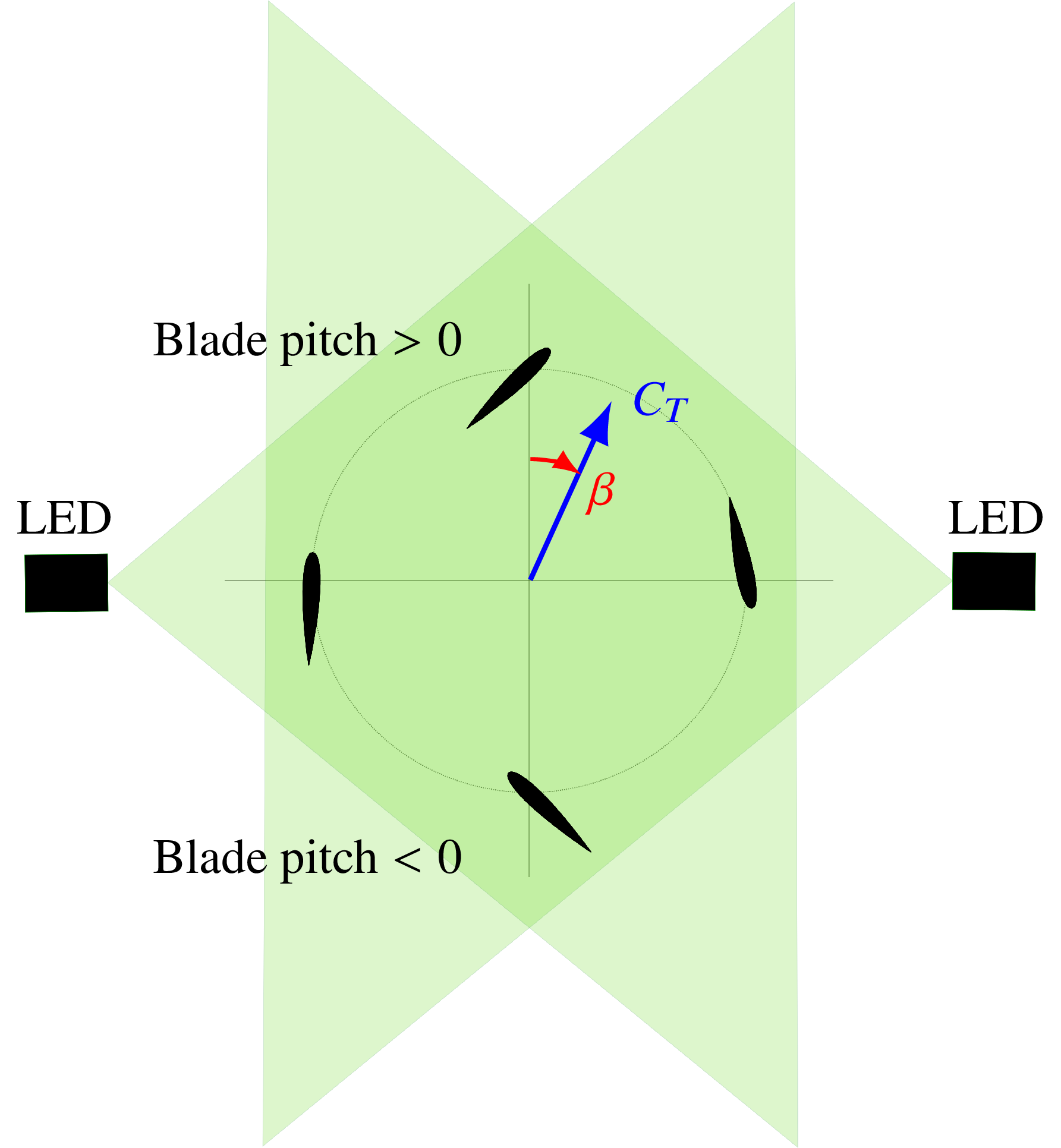}
				\caption{}
				\label{fig:PIV_setup_and_axes}
			\end{subfigure}
		\end{minipage}
		\begin{minipage}{0.49\linewidth}
			\begin{subfigure}{\linewidth}
				\centering
				\includegraphics[height=0.6\linewidth, trim=15cm 5cm 15cm 5cm,clip=true, angle=90]{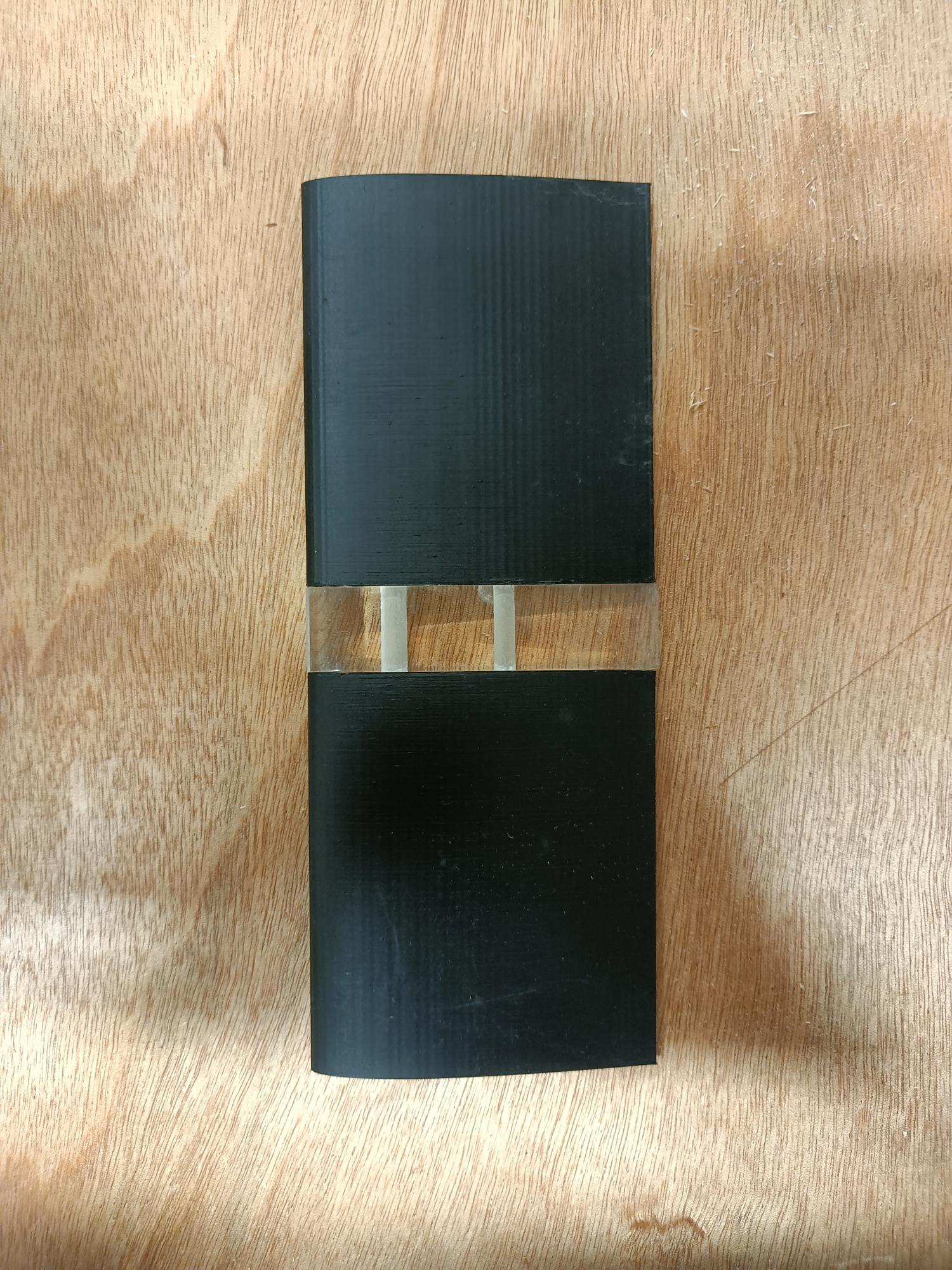}
				\caption{}
				\label{fig:PIV_setup_clear_blade}
			\end{subfigure}
			\begin{subfigure}{\linewidth}
				\centering
				\includegraphics[width=0.45\linewidth,trim=0cm 25cm 0cm 0cm,clip=true, angle=180]{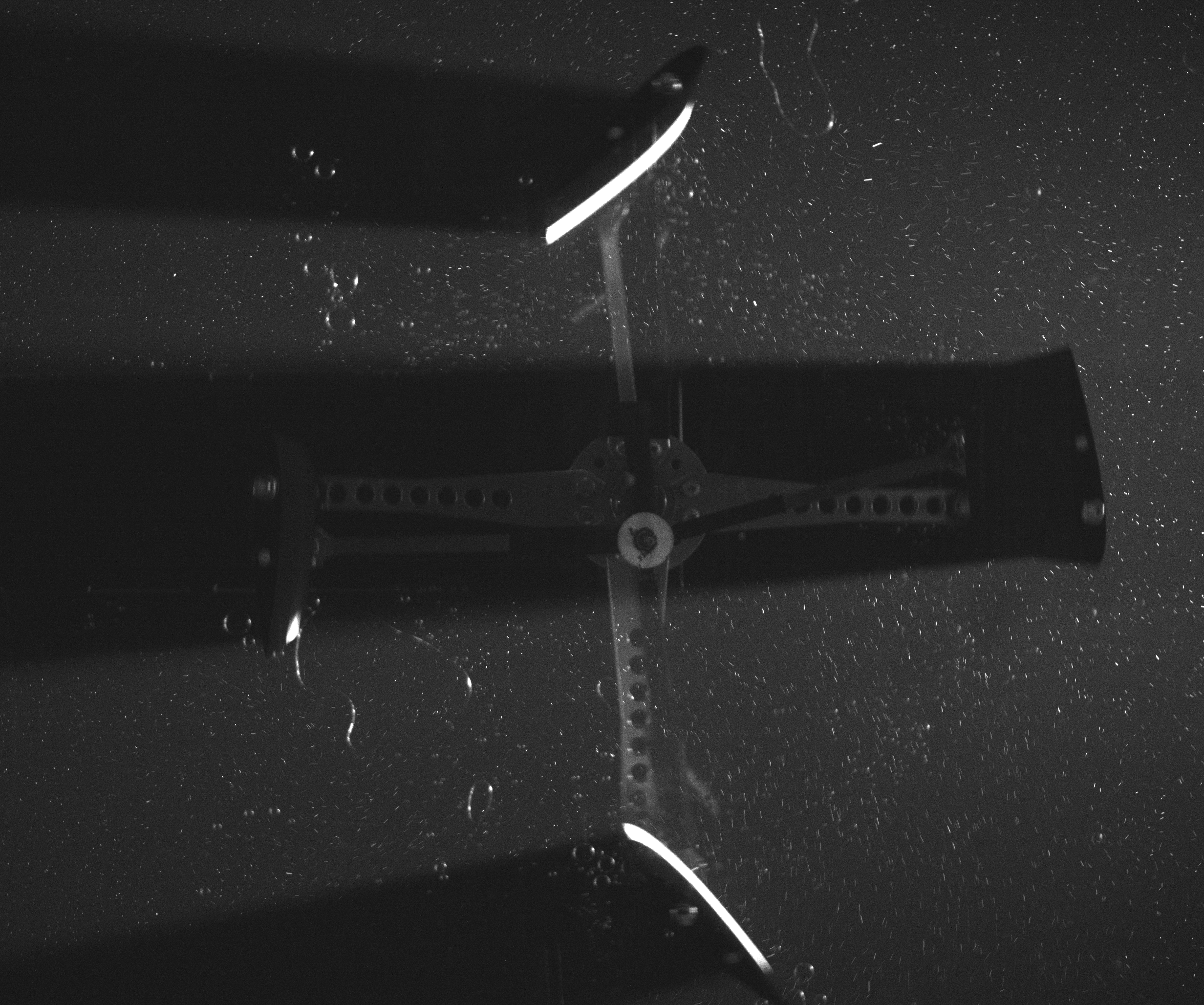}
				\caption{}
			\end{subfigure}
			\begin{subfigure}{\linewidth}
				\centering
				\includegraphics[width=0.45\linewidth]{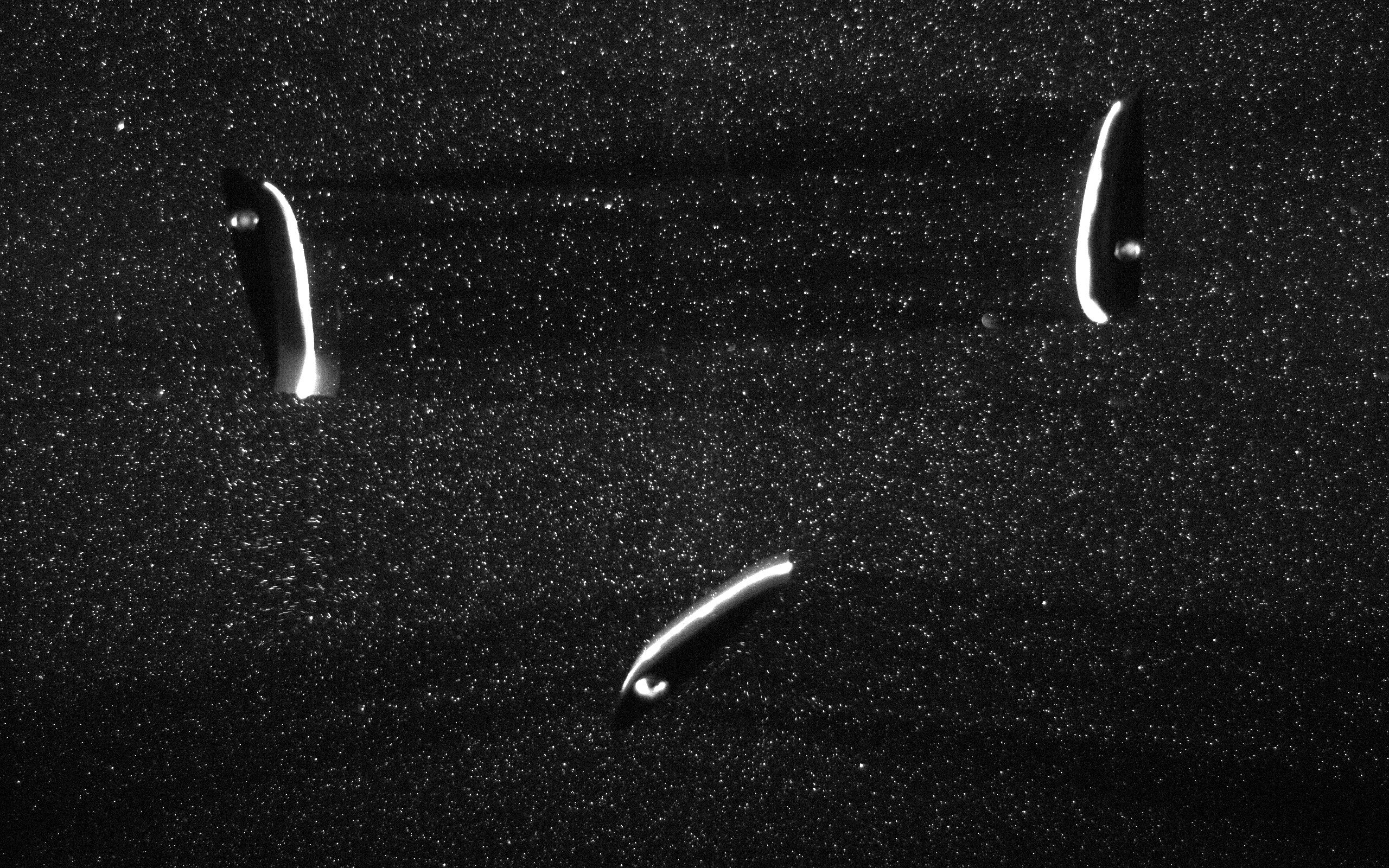}
				\caption{}
			\end{subfigure}
		\end{minipage}
		
		\caption{PIV setup using (a) 2 LED light sources. $C_T$ is the resultant coefficient of thrust and $\beta$ is the thrust vector angle. (b) Blades with clear acrylic mid-section. (c) Shadow regions when using a single light source and fully opaque blades. (d) Minimal shadow when using two light sources and blades with a transparent mid-section.}
		\label{fig:PIV_setup}
	\end{figure}
	
	To verify the final results of the optimisation process, the baseline NACA 0015 and the optimised aerofoil were tested experimentally. To aid explaining the results, the flowfields were determined with particle image velocimetry (PIV). The study was done in the University of Southampton's recirculating water tunnel featuring a test section of $1.2\, \unit{\metre}\times0.9\, \unit{\metre}\times8\, \unit{\metre}$. The tunnel was water-filled but operated without freestream flow to replicate hover conditions.

	The experiments were conducted using a modular cyclorotor rig developed and constructed at the University of Southampton (Fig. \ref{fig:experimental_setup}). The rig is a modular design with 1-4 detachable arms and blades. The setup was suspended from an ATI Delta SI-660-60 force sensor placed such that its axis was inline with the rotation axis of the rotor. This sensor has an uncertainty of $\pm 0.125\, \unit{\newton}$ in $x$ and $y$ force and $\pm\num{7.5e-3}\, \unit{\newton\metre}$. This uncertainty along with 1 standard deviation across repeated measurements is represented as a shaded area in Figure \ref{fig:forces_single_blade}. Forces were measured first in air to determine non-hydrodynamic loads associated with the rotating mechanism, including inertial loads due to blade pitching, off-centre mass effects, bearing/frictional losses and any mechanical imbalance. These phase-dependent loads were then subtracted from the corresponding measurements obtained with the rotor operating in water, isolating the hydrodynamic loads. A sampling frequency of 1000Hz was used to acquire the data. To ensure that the blade aerodynamics has reached a statistically stationary condition, forces were recorded for 120 blade-passes. This was experimentally found to be sufficient for 10-cycle mean forces/torques to converge. From measurements both with and without rotor motion, the frequencies associated with hydrodynamic loads were identified as being below 5 \unit{\hertz} ($12.5 f_{\text{rotation}}$). All frequencies above this were considered noise and filtered using low-pass filter with a cutoff frequency of 5 \unit{\hertz}. Forwards-backwards filtering was applied to minimise the phase-distortion introduced by the filtering process.

	Along with the forces, flowfields were measured using particle image velocimetry (PIV). The flow was seeded with 55 \unit[]{\micro\metre} polyamide which were illuminated at the blade mid-span using overlapping LED light sources (ILA5150 GmbH LED V3 ($\lambda \approx 520\, \unit[]{\nano \metre}$)) from opposite sides of the rotor (Fig. \ref{fig:PIV_setup_and_axes}). Two Phantom V641 high-speed cameras (sensor size: 2560px $\times$ 1600px) were mounted beneath the tunnel, viewing the illuminated plane through the tunnel's glass floor. To further reduce the number of shadow regions for the 4-bladed configuration, a 2 \unit[]{\centi \metre} span section of the blades at the mid-span was replaced with clear acrylic.  These sections were cut to the correct aerofoil shape via a waterjet cutter and then hand-polished (Fig. \ref{fig:PIV_setup_clear_blade}).
	
	Image pairs were acquired at a frequency of 100 \unit{\hertz} with a time separation of $\Delta t=2500\, \unit{\micro\second}$. 250 frame-pairs were acquired continuously to capture a single revolution of the cyclorotor. The images were both acquired and processed to produce vector flowfields using LaVision's DaVis software, employing a multi-pass method with a final window size of 32 px $\times$ 32 px and 75\% overlap. DaVis was further used to stitch the vector fields from each camera to provide an overall field-of-view of $6c \times 7c$ with a resolution of 1.6 \unit{\milli\metre}. The polynomial calibration necessary to achieve this was done using a square dotboard with 10 \unit{\milli\metre} separation inside the rotor where it was visible to both cameras.
	
	To reduce the impact of erroneous vectors and small-scale turbulent features, the vector flow-fields were phase-averaged across 20 revolutions. The image acquisition was triggered at a common point within the cycle using a hall-effect sensor and a magnet attached to the rotor shaft. This ensured that all sets of 250 image pairs start from a common point within a cyclorotor's operation and thus allowed the vectors to be averaged across all sets.

	\section{Results} \label{section:results}
	
	\subsection{Optimal aerofoil for the 4-bladed cyclorotor configuration} \label{section:optimised_aerofoil}
	
	\begin{figure}[h]
		\centering
		\includegraphics{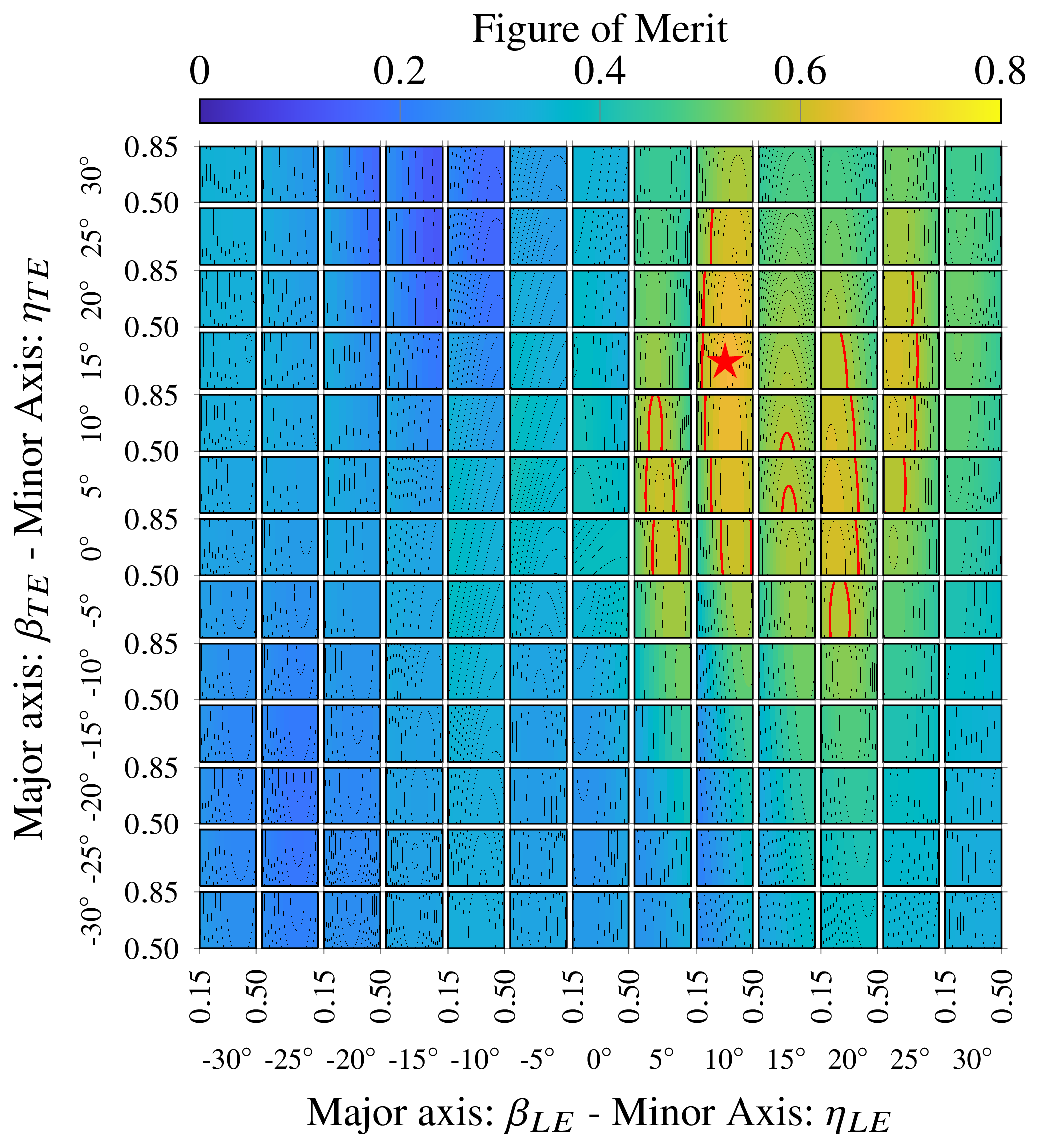}
		\caption{FM for a 4-bladed cyclorotor trends with variations in leading edge (x axis) and trailing edge (y axis) shape. Major and minor axes refer to variables represented by each individual tile and within each tile respectfully. Red star indicates approximate location of optimum. Contour lines added for easier visualisation of gradients with red contours indicating where $FM=0.9 FM_{max}$.}
		\label{fig:design_space}
	\end{figure}
	\begin{figure}[h!]
		\centering
		\includegraphics[width=0.4\linewidth]{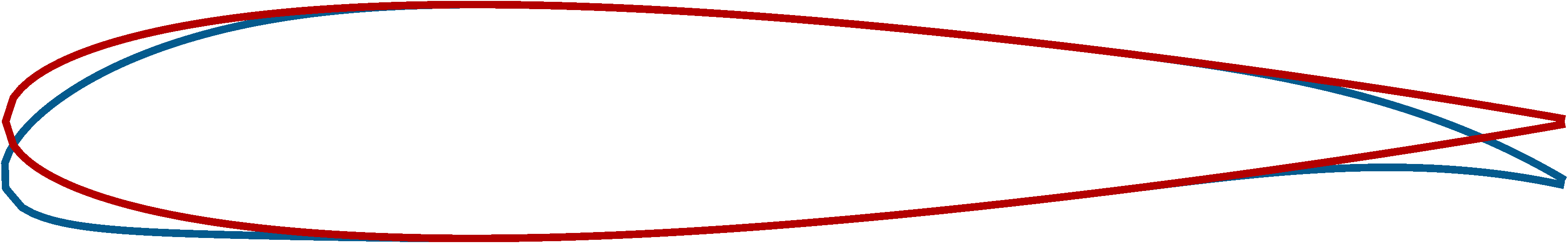}
		\caption{Optimised aerofoil (blue) for 4-bladed cyclorotor which achieved $FM=0.65$ (URANS). Original NACA 0015 in red}
		\label{fig:optimal_shape}
	\end{figure}

	The surrogate model developed for aerofoil optimisation of the four-bladed cyclorotor enables visualisation of global performance trends across the design space (Fig. \ref{fig:design_space}). Each tile within the figure represents a different combination of $\beta_{\text{LE}}$ and $\beta_{\text{TE}}$, as shown by the major axis. Within each tile, Figure of Merit is mapped as a function of hinge-point locations $\eta_{\text{LE}}$ and $\eta_{\text{TE}}$. The performance of the aerofoils exhibit strong sensitivity to the droop angles $\beta_{\text{LE}}$ and $\beta_{\text{TE}}$, while demonstrating comparatively weak dependence on the position of the hinge points.

	The optimal aerofoil geometry identified through the optimisation process exhibits a slight positive camber, with both the leading and trailing edges drooping by $\approx7 \unit{\degree}$, thus reinforcing the virtual camber effect (Fig. \ref{fig:optimal_shape}). These trends agree with the observations of previous studies on aerofoil camber in cyclorotors \citep{Tang2017UnsteadyModel, Zhang2018ThePropellers}. The design space indicates that the optimum is relatively insensitive to small geometric perturbations. The contours highlighted in red in Fig. \ref{fig:design_space} denote designs for which $FM$ is greater than or equal to 90\% of $FM_{max}$. This near-optimal region extends over a finite range of leading-edge and trailing-edge droop parameters, indicating that the selected design is not an isolated sharp optimum. From a practical design perspective, this suggests that modest manufacturing or geometric variations around the selected aerofoil would be unlikely to substantially reduce the predicted performance.

	These findings appear, at first glance, to contrast with results reported in  studies examining asymmetric pitching profiles in cyclorotors, which demonstrated that introducing a positive bias to the pitching profile could improve performance by counteracting the virtual camber's negative bias \citep{Benedict2010FundamentalApplications, Shi2022AnalysisRatio, Benedict2016EffectsCycloidal-rotor, Walther2019SymmetricNumbers}. It was therefore expected that aerofoil optimisation could exhibit a similar trend. Instead, the optimised geometry appears to further augment the virtual camber effect.

	Repeating the optimisation process for 1-3 blades resulted in very similar aerofoils, with the mean difference between aerofoil surface points being $<1\%$ of the chord-length. The small differences arise from the same low sensitivity in $FM$ to small variations in aerofoil shape observed near the optimum for the 4-bladed configuration (Fig. \ref{fig:design_space}). To reduce experimental costs, the following experimental results in this study have used the aerofoil optimised for the 4-bladed configuration.

	\subsection{LEV-dynamics of a single blade in curvilinear flow}
	
	\begin{figure}[h]
		\centering
		\includegraphics[scale=0.8]{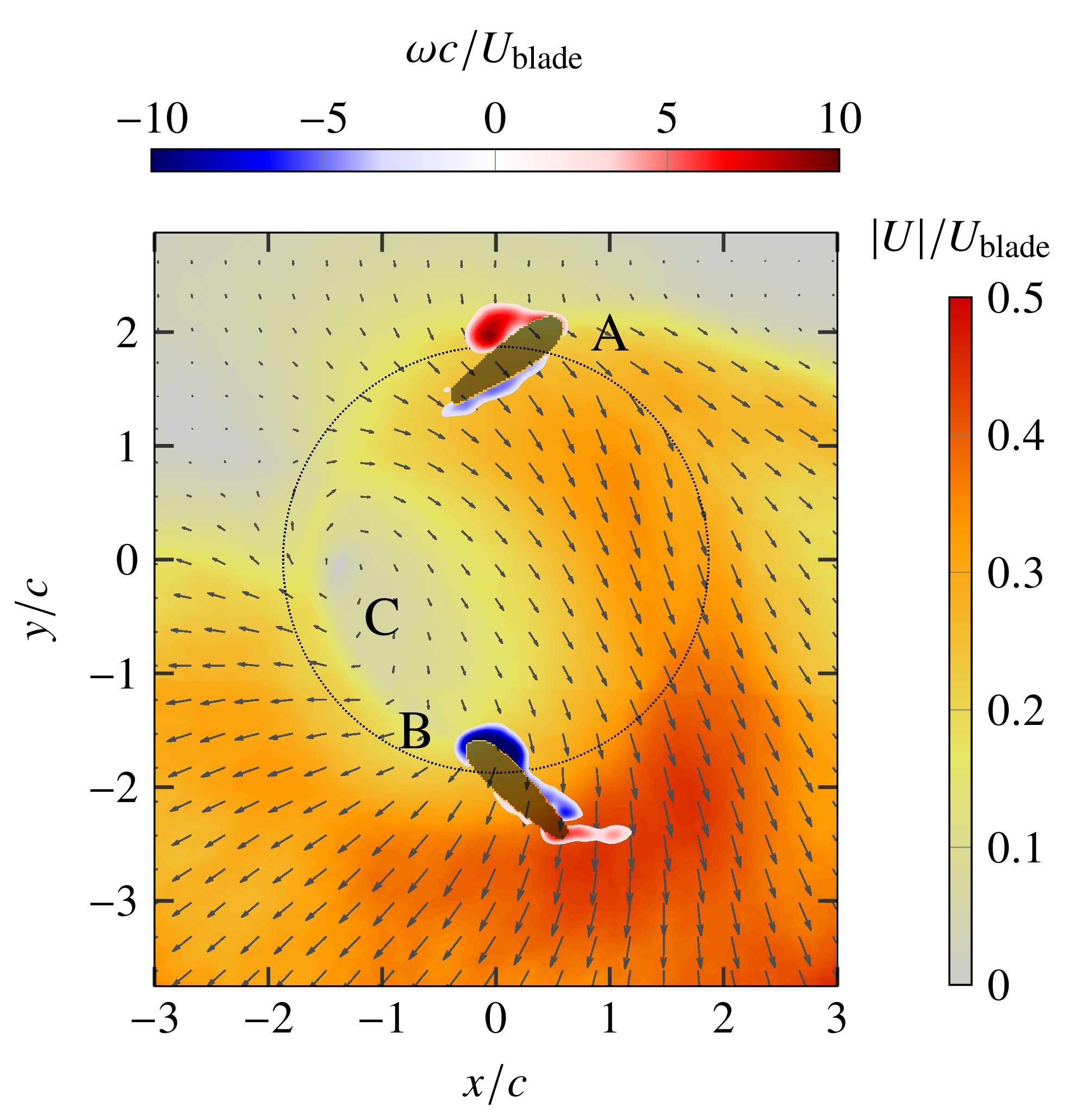}
		\caption{Mean velocity flowfield (grey-orange) overlayed with instantaneous near-blade vorticity (blue-white-red) for a single bladed cyclorotor rotating at 24 RPM. The three main features identified are (A.) The LEV generated at a positive blade pitch  $\alpha= 45\unit[]{\degree}$ and an azimuthal angle of $\psi = 270\unit{\degree}$ (B.) The LEV generated at a negative blade pitch of $\alpha= -45\unit[]{\degree}$ and an azimuthal angle of $\psi = 90\unit{\degree}$. (C) The recirculation zone.}
		\label{fig:cyclo_flow_diagram}
	\end{figure}
	\begin{figure}[h]
		\centering
		\includegraphics{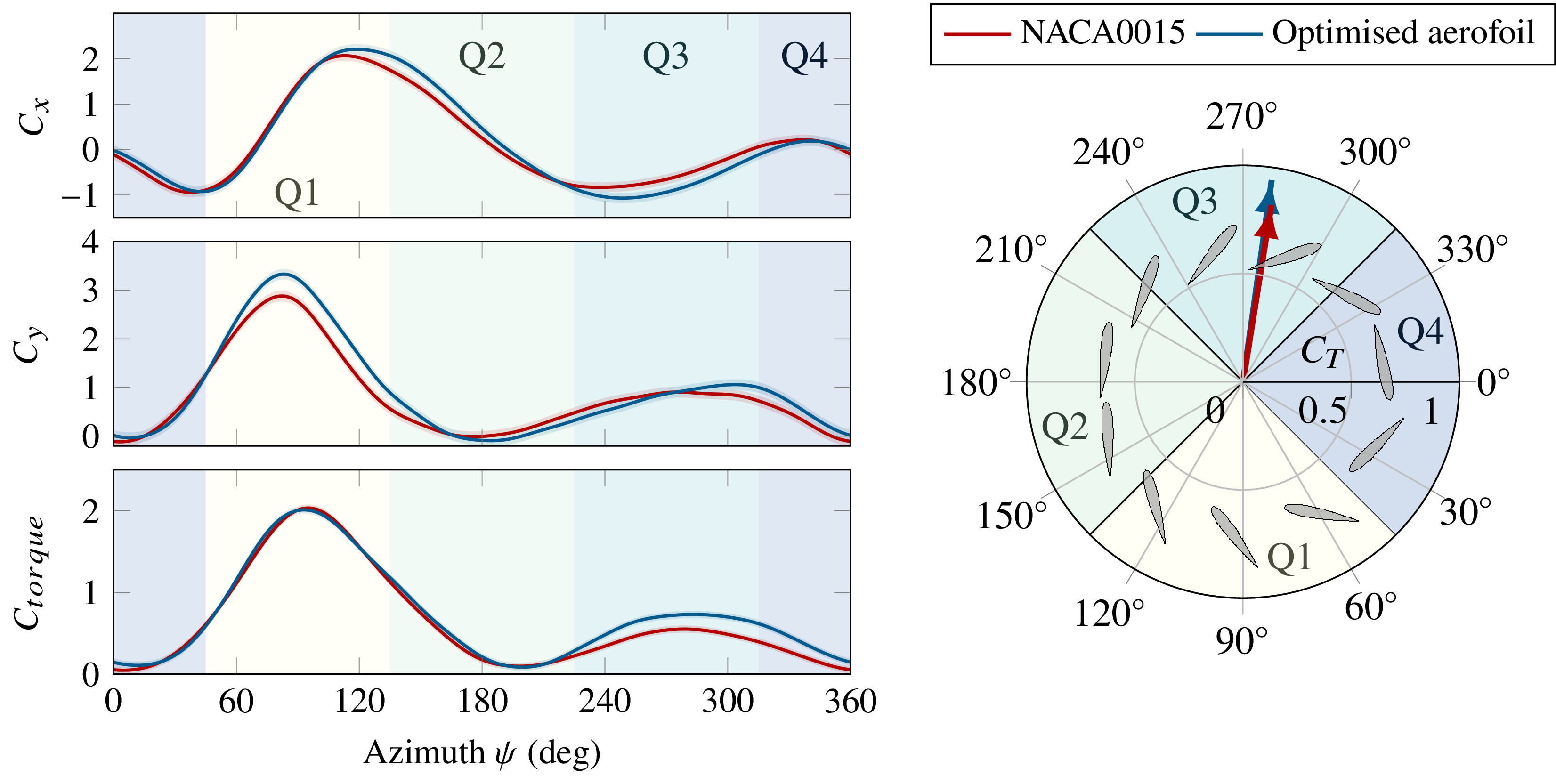}
		\caption{Force and torque coefficients over a single revolution of a 1-bladed cyclorotor using baseline (red) and optimised aerofoil (blue). Shaded regions represent overall uncertainty.}
		\label{fig:forces_single_blade}
	\end{figure}
	
	To illustrate the principal flow features, PIV data for a single-bladed cyclorotor employing the baseline NACA 0015 aerofoil are presented, showing the mean velocity field alongside instantaneous vorticity distributions around the blade at two representative azimuthal positions (Fig. \ref{fig:cyclo_flow_diagram}). The azimuthal positions are presented in Fig. \ref{fig:forces_single_blade}.  As the blade goes through its maximum (A) and minimum (B) pitch it undergoes dynamic stall and shed large LEVs (Fig. \ref{fig:cyclo_flow_diagram}). The blade kinematics induces a mean throughflow where the flow is ingested into the rotor's upper-half, passes around the right-hand side of the rotor and is then ejected out of the bottom across a wide arc. The flow is accelerated both upon entering and exiting the rotor. As the blade re-enters the upper-half of the rotor ($\psi=\ang{180}$), its motion opposes the induced throughflow and generates a recirculatory flow region (C). The LEV created during negative pitch (B) will detach and remain in this region as it breaks down and dissipates. In contrast, the LEV generated during positive pitch, typically detaches around ($\psi=\ang{300}$) and is convected rightwards, away from the rotor, by the induced mean flow.

	The force and torque profiles on a single bladed cyclorotor can be divided into four quadrants (Q1-Q4) (Fig. \ref{fig:forces_single_blade}).  The instantaneous resultant thrust ($C_T$) does not maintain a constant direction over a cycle, necessitating a component-wise analysis of $x$ and $y$ force contributions.

	The force/torque components for the baseline NACA 0015 all exhibit a primary peak in Q1, corresponding to the minimum blade pitch (-\ang{45}), and a smaller secondary peak in Q3, corresponding to the maximum blade pitch (+\ang{45}). The asymmetry in the peaks can be attributed to the virtual camber effect which introduces a negative offset to the effective angle-of-attack (Fig. \ref{fig:virtual_camber_cyclo}). This offset increases thrust production under negative blade pitch conditions (Q1) while reducing thrust generation under positive blade pitch (Q3). In Q2 and Q4, the blade pitch returns to zero, resulting in comparatively lower aerodynamic loading in these regions.

	\begin{figure}[t]
		\centering
		\includegraphics{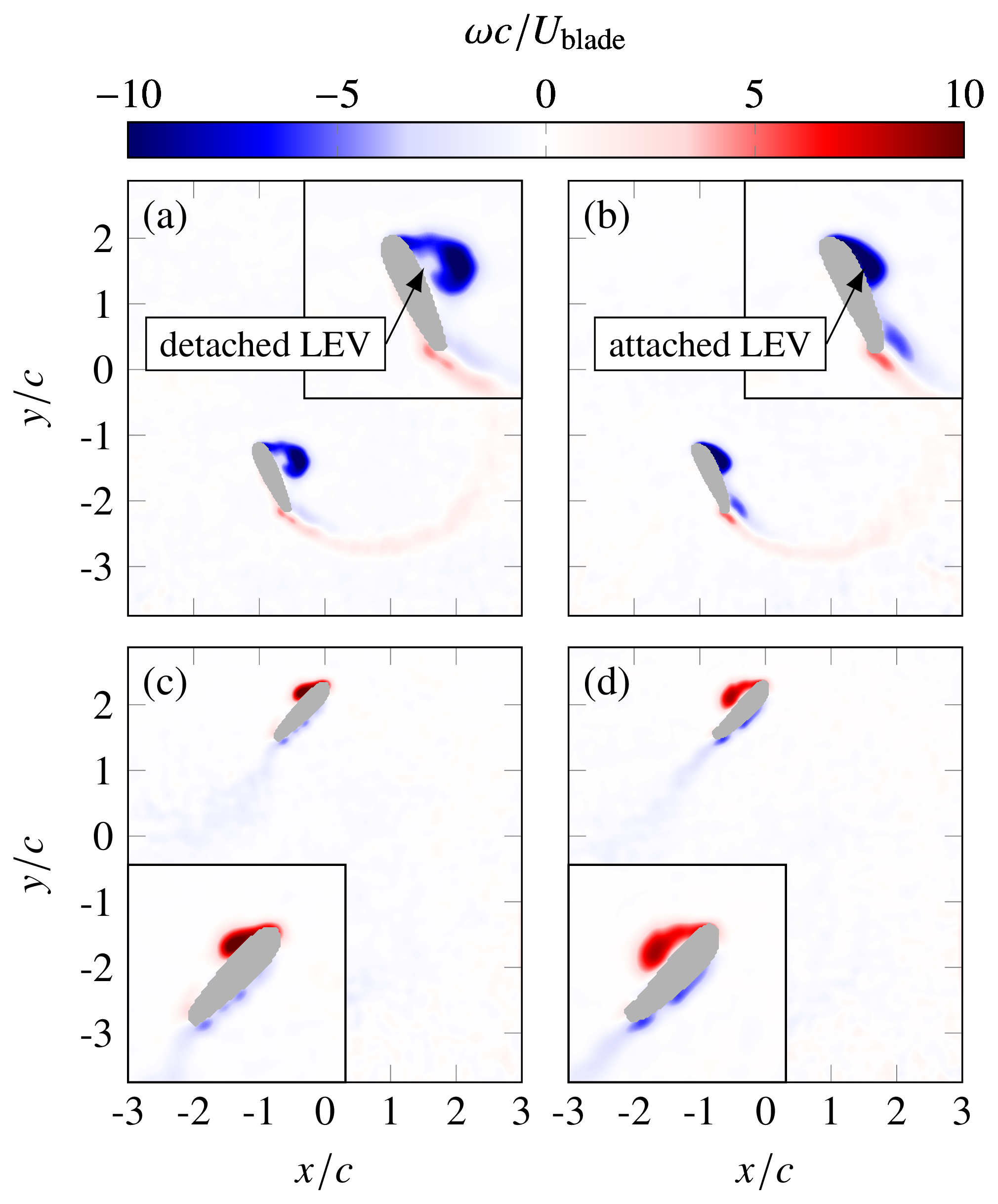}
		\caption{Differences in LEV during negative pitch when using NACA 0015 (a and c) and optimised aerofoil (b and d) for a 1-bladed configuration. Shown for blade azimuth (a and b) $\psi=119\unit{\degree}$ and (c and d) $\psi=261\unit{\degree}$}
		\label{fig:1_blade_vorticity}
		\phantomsubcaption \label{fig:1_blade_vorticity_A}
		\phantomsubcaption \label{fig:1_blade_vorticity_B}
		\phantomsubcaption \label{fig:1_blade_vorticity_C}
		\phantomsubcaption \label{fig:1_blade_vorticity_D}
	\end{figure}

	The $C_y$ and $C_{torque}$ components remain positive over the majority of the rotational cycle, with both primary and secondary peaks being positive. In contrast, the peaks of $C_x$ have opposing signs, causing them to largely cancel out when averaged over a full revolution. This results in a relatively small mean value for $C_x$ compared to $C_y$, despite their similar instantaneous magnitudes. The greater magnitude of the primary peak due to virtual camber makes the mean $C_x$ value positive, causing the mean resultant force vector $C_T$ to rotate $21\unit{\degree}$ clockwise from the vertical ($\beta$).

	When using the optimised aerofoil of the 4-bladed configuration (Fig. \ref{fig:optimal_shape}), the primary differences seen are a 16\% increase in $C_y$ during the primary peak (Q1) and a 33\% increase in $C_{torque}$ during the secondary peak (Q3). Otherwise however, the same general trends of the forces and torques remain consistent with those observed for the baseline NACA 0015.
	
	To explain these observations, the vorticity fields derived from PIV data are examined. For both aerofoil types, LEVs are shed as the blades undergo dynamic stall at minimum pitch (Fig. \ref{fig:1_blade_vorticity_A}-\ref{fig:1_blade_vorticity_B}) and maximum pitch (Fig. \ref{fig:1_blade_vorticity_C}-\ref{fig:1_blade_vorticity_D}). Due to the negative effective angle-of-attack offset induced by virtual camber, the absolute magnitude of angle-of-attack is greater during the negative pitch phase. Consequently, a more prominent negative LEV is observed at minimum pitch (Fig. \ref{fig:1_blade_vorticity_A}-\ref{fig:1_blade_vorticity_B}). For the optimised aerofoil, the negative LEV is much smaller and more closely attached to the surface of the blade (Fig. \ref{fig:1_blade_vorticity_B}). Conversely the positive LEV during positive pitch shows greater separation from the blade surface (Fig. \ref{fig:1_blade_vorticity_D}). Since flow separation is typically accompanied by a loss in lift and increase in drag, these differences in the LEVs align with what was observed for the single blade forces, which showed the optimised aerofoil improving thrust production in Q1 but increasing torque in Q3 (Fig \ref{fig:forces_single_blade}).

	\begin{figure}[t]
		\centering
		\includegraphics{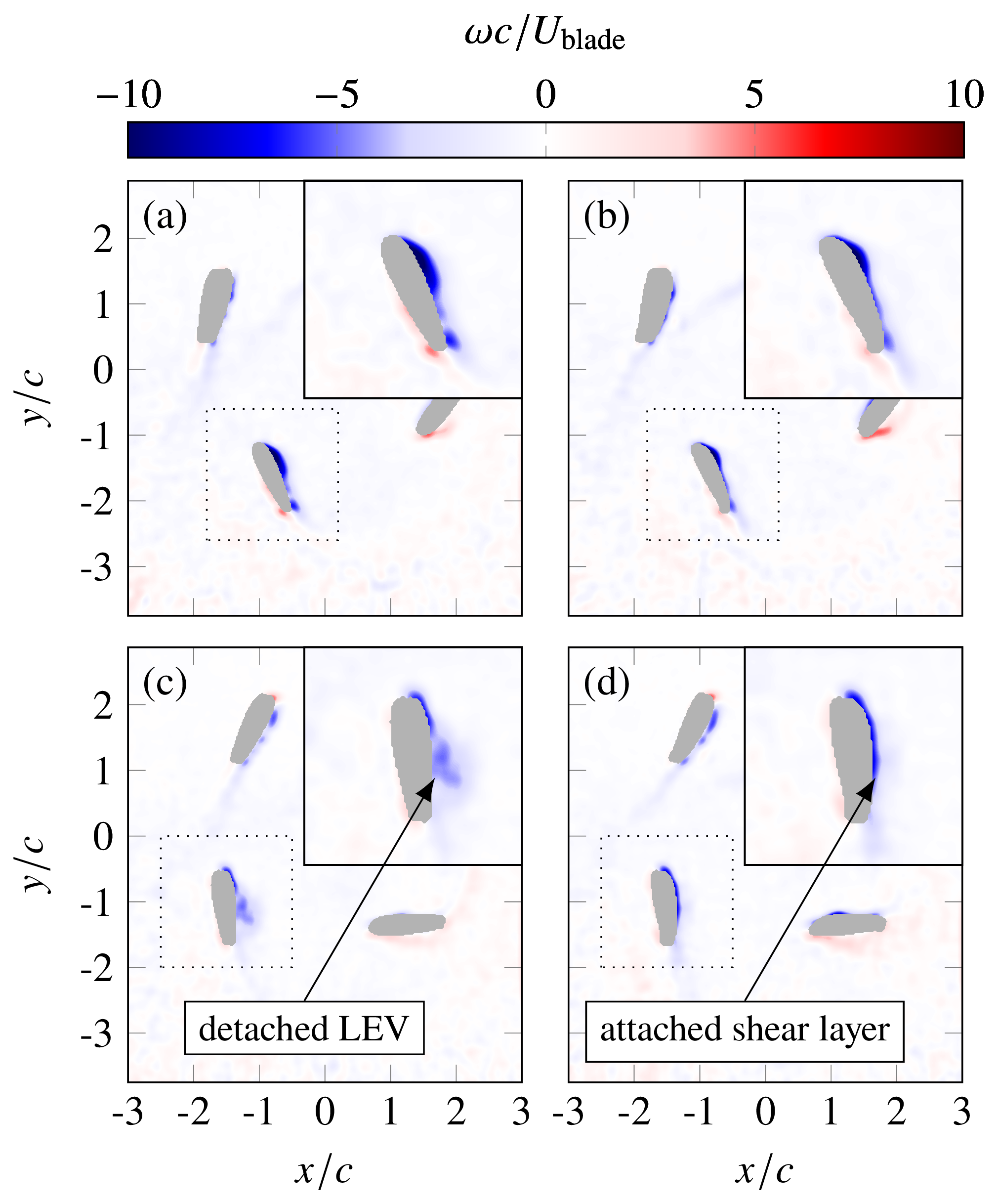}
		\caption{Differences in LEV during negative pitch when using NACA 0015 (a and c) and optimised aerofoil (b and d) for a 4-bladed cyclorotor. Shown for blade azimuth (a and b) $\psi=119\unit{\degree}$ and (c and d) $\psi=147\unit{\degree}$}
		\label{fig:4_blade_vorticity}
		\phantomsubcaption \label{fig:4_blade_vorticity_A}
		\phantomsubcaption \label{fig:4_blade_vorticity_B}
		\phantomsubcaption \label{fig:4_blade_vorticity_C}
		\phantomsubcaption \label{fig:4_blade_vorticity_D}
	\end{figure}

	\subsection{Impact of blade-count upon LEV characteristics} \label{section:4_blade_rotor}

	Differences in the LEV shedding characteristics are observed with increasing blade count from 1 to 4 despite the blade kinematics remaining constant. In the single-bladed baseline case, the negative LEV is characterised by its larger size and pronounced separation (Fig. \ref{fig:1_blade_vorticity_A}). In the 4-bladed configuration with the same baseline aerofoil however, the same LEV appears to be much smaller and less distinct (Fig. \ref{fig:4_blade_vorticity_A}). While the LEV does eventually detach, separation occurs much later in the cycle and after the primary thrust peak has ended (Fig. \ref{fig:4_blade_vorticity_B}). The relative differences between the baseline and optimised aerofoil for the 4-bladed configuration do however remain similar to what was seen for the 1-bladed configuration. With the optimised aerofoil, the negative LEV is further reduced in size and its separation is suppressed (Fig. \ref{fig:4_blade_vorticity_C},\ref{fig:4_blade_vorticity_D}).

\begin{figure}[h]
		\centering
		\includegraphics[scale=0.9]{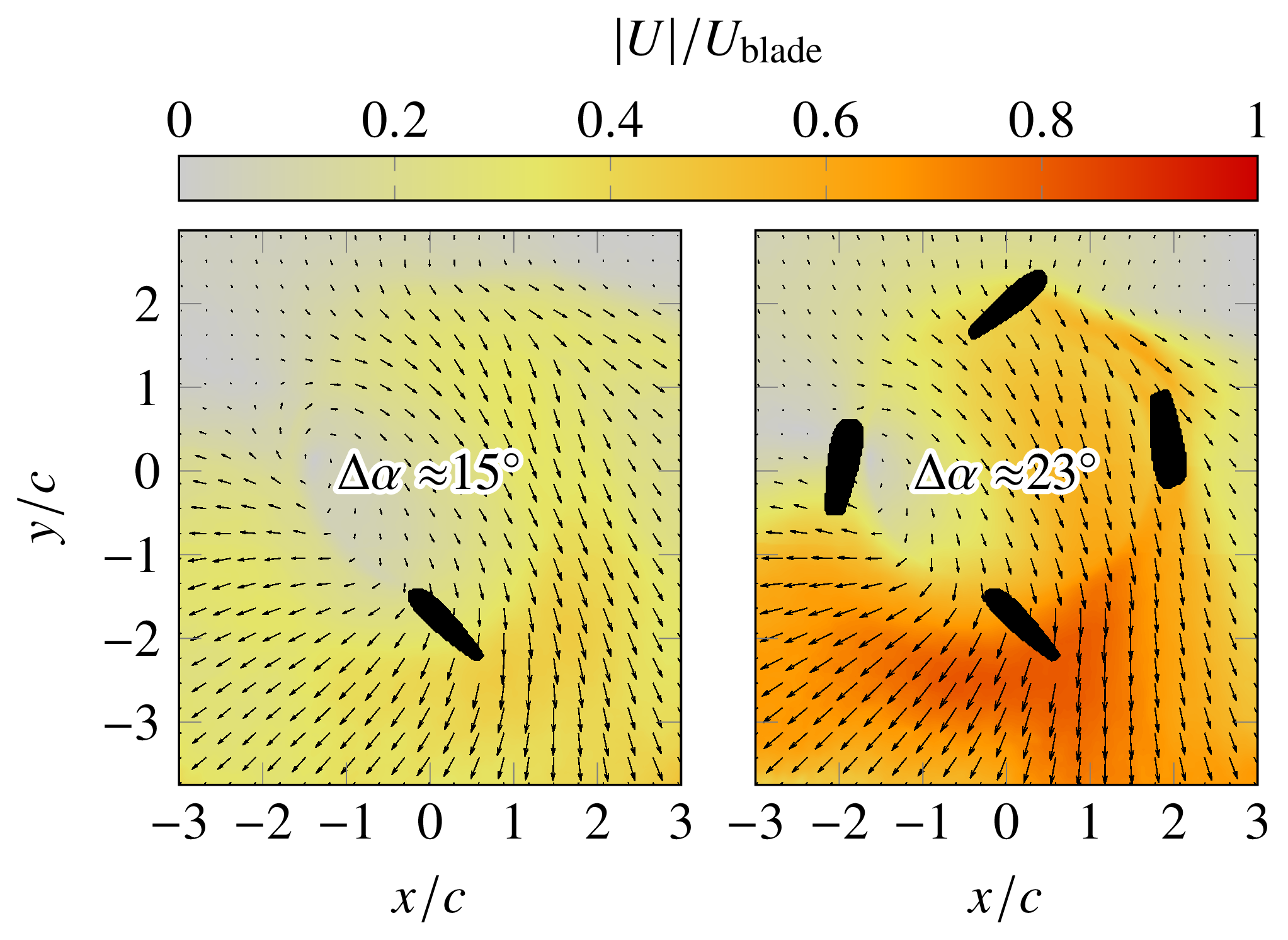}
		\caption{Mean flowfield for 1- (left) and 4- (right) bladed cyclorotor using NACA 0015 aerofoil alongside estimate for change in effective angle-of-attack imparted by throughflow.}
		\label{fig:mean_flow}
	\end{figure}

	This change in the LEV characteristics with different blade-counts can be explained through an examination of the mean flowfields (Fig. \ref{fig:mean_flow}). For both the one- and four-bladed configurations, the overall mean-flow structure remains consistent: flow is ingested through the top, passed around one side of the rotor, and is then ejected radially across a wide arc. On the opposite side, where the blade motion is directed against the induced throughflow, a recirculating-flow region develops. The primary distinction lies in the magnitude of this throughflow, with the maximum mean throughflow speed ($U_t$) increasing by $\approx85\%$ in the four-bladed configuration relative to the single-bladed case).

	Through conservation of momentum, the magnitude of the throughflow in a cyclorotor is approximately proportional to the square-root of its thrust ($T$),
	\begin{equation}
		U_t \propto \sqrt{T},
		\label{eq:Ut_thrust}
	\end{equation}
	and for a fixed pitching profile the thrust is proportional to both the blade-count ($N_b$) and the blade speed ($U_b$),
	\begin{equation}
		T \propto N_b U_b^2
	\end{equation}
	
\begin{figure}[t]
		\centering
		\includegraphics[scale=0.9]{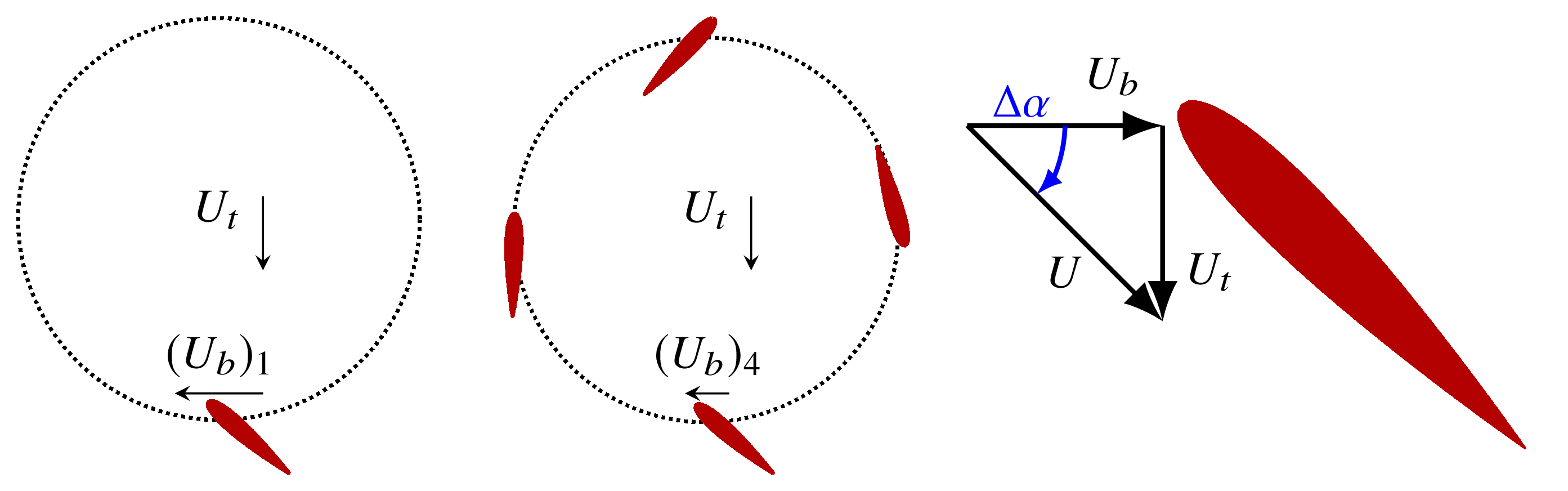}
		\caption{Impact of blade-count on downwash angle for fixed thrust/throughflow}
		\label{fig:downwash_maths}
	\end{figure}

	Consequently when comparing two cyclorotors with different blade-counts for either a fixed RPM or fixed thrust (and therefore throughflow):
	\begin{equation}
		\left( \frac{U_t}{U_b}\right)_{N_b = 1} \quad < \quad \left( \frac{U_t}{U_b}\right)_{N_b=4} 
	\end{equation}
	This ratio is important, since the throughflow suppresses the angle-of-attack of the blades in the lower half of the rotor where the majority of the thrust is produced. The extent to which the angle-of-attack is suppressed is then a function of the ratio of throughflow to blade-speed (Fig. \ref{fig:downwash_maths}) and can be mathematically represented by:
	\begin{equation}
		\Delta\alpha=\arctan{\left( \frac{U_t}{U_b}\right) }
	\end{equation}
	Consequently, the effective angle-of-attack of blades in a cyclorotor with a higher blade-count is significantly more suppressed. This is demonstrated Fig. \ref{fig:mean_flow}, using the average flowspeed within the rotor cage to estimate $U_t$. This difference in $\Delta\alpha$ explains the differences seen in the negative LEV when comparing the 1- and 4-bladed results (Figs. \ref{fig:1_blade_vorticity}-\ref{fig:4_blade_vorticity}). With a larger effective angle-of-attack, the blade in the 1-bladed configuration undergoes a much deeper dynamic stall, as characterised by the large LEV that is shed \citep{McCroskey1981TheStall}. Conversely, the blades in the 4-bladed configuration undergo a much lighter dynamic stall with a lower degree of flow separation.

	\subsection{Influence of LEV separation upon cyclorotor efficiency}
	
	\begin{figure}[h]
		\centering
		\includegraphics[scale=0.9]{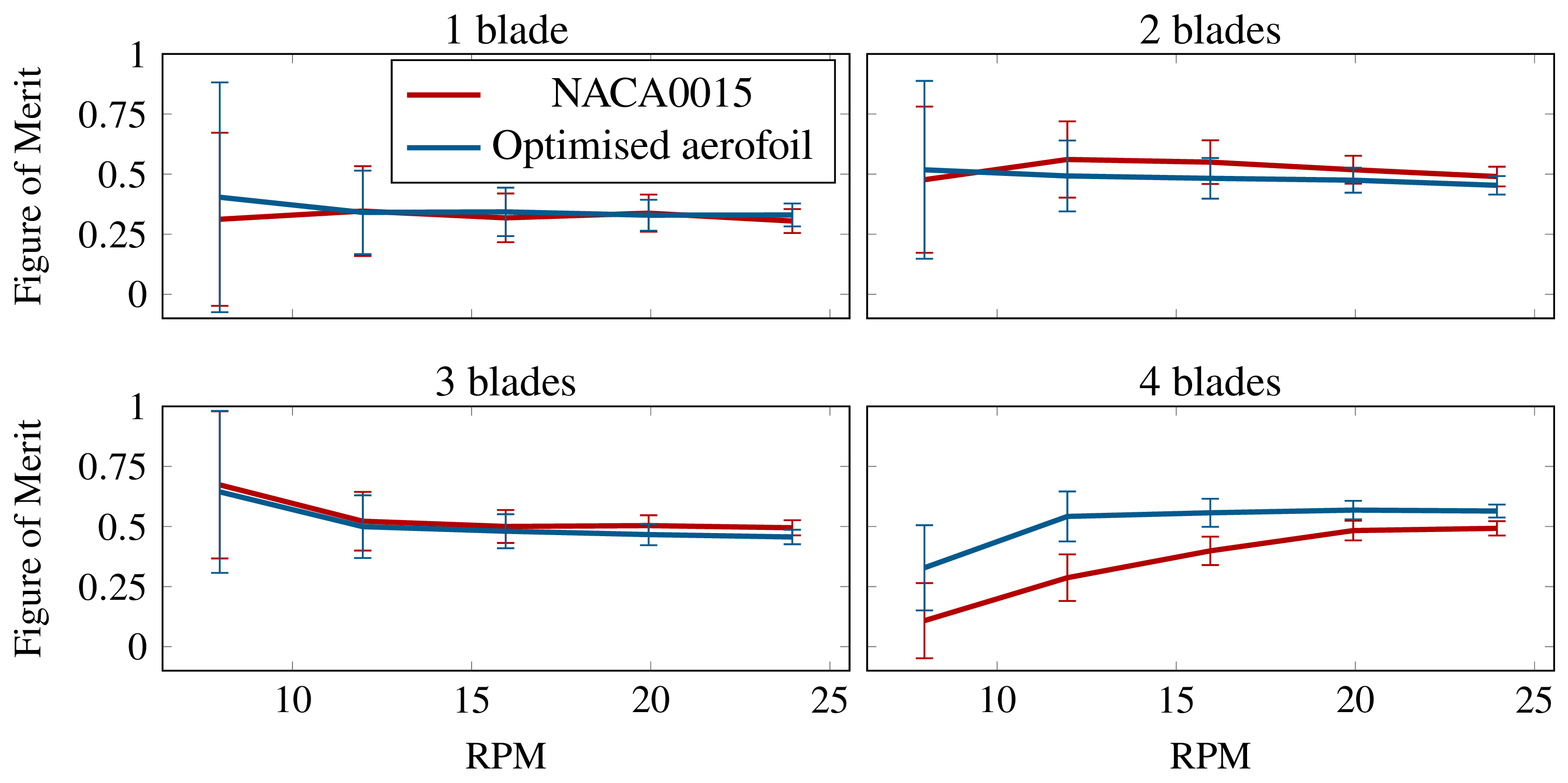}
		\caption{Figure of Merit from experimental investigation for different blade-counts when using an aerofoil optimised for the 4-bladed configuration}
		\label{fig:FM}
	\end{figure}
	
	The connection between blade-count and the extremity of dynamic stall has an impact upon aerofoil optimisation. Using the aerofoil optimised for the 4-bladed configuration, Figure of Merit was measured experimentally for 1-4 blades across a range of RPM (Fig. \ref{fig:FM}). For the 4-bladed configuration that the aerofoil was optimised for, a notable improvement can be seen when using the optimised aerofoil (14\% at 24 RPM). For all blade-counts lower than 4, however, there is a negligible difference to performance when using the optimised aerofoil. CFD showed that using the true optimised aerofoil for each blade-count did not improve $FM$ either, due to the low sensitivity to aerofoil geometry around the optimum (Section \ref{section:optimised_aerofoil}).

	The observed trends are attributable to differences in extremity of the dynamic stall noted during the primary thrust peak when using different blade-counts (Sec. \ref{section:4_blade_rotor}). For the rotor geometry used in this particular study, the dynamic stall for blade-counts less than 4 is likely too deep for any change in aerofoil shape to prevent the LEV from separating. This can be seen in the single-bladed results, where the LEV during the primary thrust peak is very large and distinct (Fig. \ref{fig:1_blade_vorticity_A}). For 4 blades however, the dynamic stall was much lighter with a LEV that separates much later (Fig. \ref{fig:4_blade_vorticity}). This consequently allows an adjustment in aerofoil shape to suppress the LEV. Given the lift degradation and drag augmentation associated with separated flow, this distinction explains why measurable performance improvements emerge in the 4-bladed configuration but remain absent in the lower blade-counts. A similar observation has been made for VATs where vortex generators were used to improve performance by preventing flow separation \citep{DeTavernier2021ControllingAirfoil}.

	\subsection{The generality of results for other configurations}
	In this study, a strong dependency has been demonstrated between cyclorotor performance and the behaviour of dynamic stall LEVs. Consequently, while this study has primarily focused on the relationship between these LEVs and rotor blade count, the ultimate efficacy of aerofoil shape optimisation will also be dependent on all operational and geometric parameters that dictate dynamic stall behaviour.
	
	For instance, the laminar boundary layers characteristic of the low-Reynolds-number regime studied here are exceptionally sensitive to adverse pressure gradients. At higher Reynolds numbers however, the transition to turbulent boundary layers can significantly delay flow separation \citep{Winslow2018Basic104105, Katz2001LowMethods}. Similarly, surface roughness profiles and their associated boundary layer perturbations will alter the onset and progression of dynamic stall \citep{Shum2025EffectMotion}. Furthermore, modifications to the rotor geometry that alter the chord-to-radius ($c/R$) ratio will also adjust the reduced frequency of the system, thereby altering the aerodynamic unsteadiness and thus the dynamic stall behaviour \citep{McCroskey1981TheStall}. The influence of these parameters upon dynamic stall were the primary reason why solidity in this study is controlled via blade-count. Doing so allows changes in solidity without needing to consider the impact of these other effects.
	
	The core conclusion of the present work is ultimately mechanistic rather than geometric: aerofoil shape optimisation is effective when the operational regime allows the geometric modifications to successfully suppress or mitigate LEV separation. While the specific optimum profile is inherently configuration-dependent, the underlying design objective—controlling LEV attachment—is expected to remain a universal principle for cyclorotor design and may offer valuable insights for other curvilinear dynamic stall applications.
	
	To further contextualise this point, it has been observed in vertical-axis wind turbines (VAWTs) that aerodynamic performance is enhanced when the blades operate within light, rather than deep, dynamic stall regimes \citep{LeFouest2022TheTurbines}. Despite fundamental differences in global kinematics—most notably the presence of a freestream inflow velocity in VAWTs compared to the quiescent hover conditions investigated here—these rotating systems exhibit the same trends with respect to dynamic stall extremity. This agreement supports the assertion that managing LEV development and avoiding the deep dynamic stall regime, is a broadly applicable aerodynamic principle rather than a phenomenon isolated to a specific cyclorotor configuration.
	
	\section{Conclusion}
	
	The present study investigated the key physical mechanisms influencing the optimisation of foils in curvilinear flow. URANS simulations in conjunction with a Kriging model were used to optimise the aerofoil of a 4-bladed cyclorotor. The final optimised aerofoil was then experimentally assessed with a physical experiment measuring both the forces and the flowfields.
	
	Aerofoil geometry optimisation was shown to improve cyclorotor performance. For the four-bladed configuration, the optimal aerofoil—characterised by approximately $7\unit{\degree}$ leading- and trailing-edge droop—suppressed separation of the leading-edge vortex formed during the primary thrust peak. Mitigation of separation-associated lift loss and drag augmentation enabled a 14\% increase in Figure of Merit.
	
	This performance enhancement is subject to a governing constraint: optimisation effectiveness depends strongly on dynamic stall severity, which is regulated by rotor solidity. Under deep dynamic stall conditions, leading-edge vortex separation remains dominant and cannot be suppressed through aerofoil shape modification alone. Effective optimisation therefore requires a flow regime in which the vortex remains predominantly attached prior to geometric intervention.

	For cyclorotors operating in hover, this dependency renders optimisation ineffective at low solidity. Higher-solidity configurations induce stronger throughflow, reducing effective blade incidence during the primary thrust phase and promoting the lighter dynamic stall conditions necessary for vortex attachment and optimisation viability.

	Overall, aerofoil optimisation in cyclorotors cannot be conducted in isolation. Rotor solidity, reduced frequency, and blade kinematics collectively govern dynamic stall behaviour and must be considered concurrently with aerofoil geometry in defining optimal rotor design.

	\appendix
	
	\section{Limitations of URANS simulation} \label{appendix:URANS}
	
	While the present 2D URANS methodology is commonly used for curvilinear flow problems, its limitations should be acknowledged. In a study comparing 2D and 3D URANS simulations of a cyclorotor, it has been noted that 2D simulations can overestimate the strength of the LEVs shed during dynamic stall, partly because they do not capture finite-span effects such as tip-vortex induced downwash \citep{Yu2016Two-dimensionalHover}. In addition, the mechanisms governing LEV decay/breakdown and loss of coherence are intrinsically three-dimensional and therefore cannot be represented directly in a two-dimensional simulation \citep{Rivera1995TheTurbulence, Zurman-Nasution2020InfluenceFlapping}. In the URANS method used in the current study, the eddy viscosity model used is inherently dissipative, allowing the vortices to eventually decay despite the lack of three-dimensionality \citep{Pope2000TurbulentFlows}. Similarly, while the $k-\omega$ SST turbulence model is a common model for separated flows, the model assumes a fully turbulent flow and thus does not resolve laminar-turbulent transition unless coupled with a transition model \citep{Pope2000TurbulentFlows, Wilcox2010TurbulenceCFD}. Boundary layers consequently may be treated as turbulent earlier than in the physical low-Re flow, increasing near-wall mixing and potentially delaying or suppressing separation \cite{Bangga2017DynamicModel, Dave2021ComparisonFlow}. Therefore, the simulations should be interpreted primarily as a comparative optimisation tool to identify promising aerofoil geometries, with the resulting performance trends assessed experimentally.

\begin{figure}[h]
		\centering
		\includegraphics{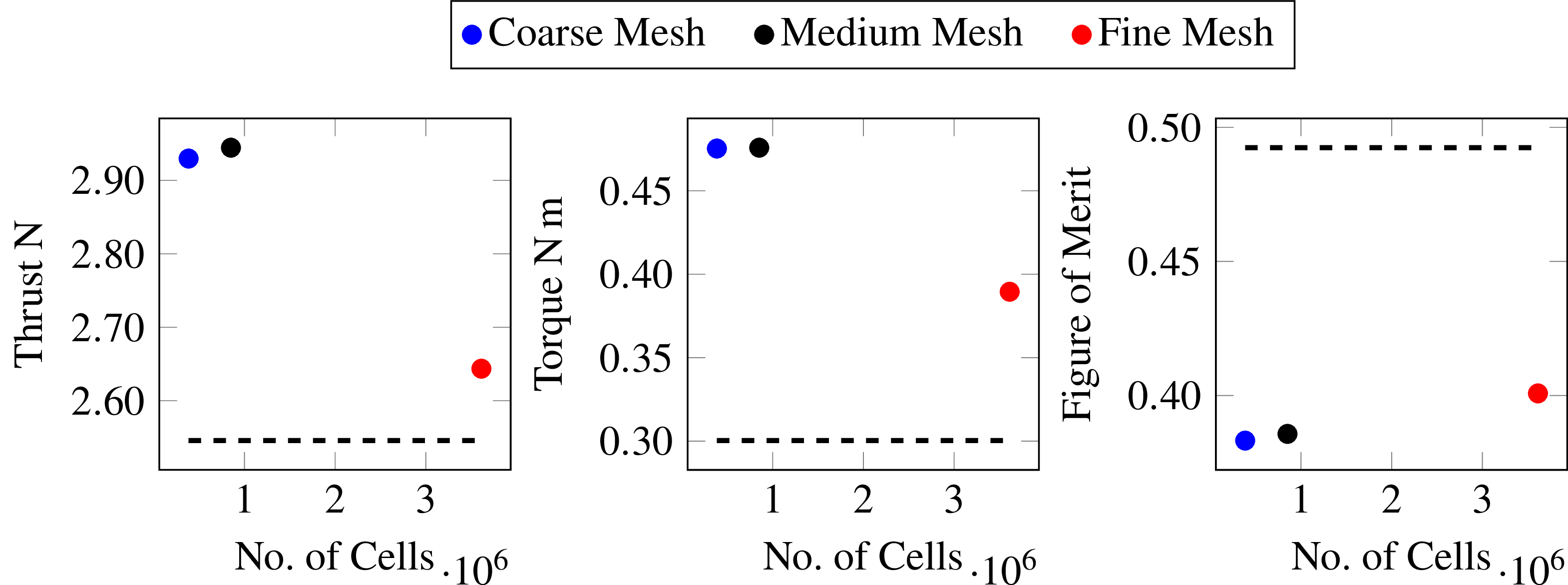}
		\caption{Mesh sensitivity for cyclorotor performance metrics. Experimentally found values shown as dashed line}
		\label{fig:meshConvergence}
	\end{figure}

	These limitations of 2D URANS also influenced the choice of mesh density. After evaluating thrust, torque, and $FM$ across three different mesh densities (400k, 800k, and 4,000k), thrust and torque were consistently over-predicted, while $FM$ was under-predicted (Fig. \ref{fig:meshConvergence}). Although the high-density mesh improved the prediction of thrust and torque, it did so at the expense of much larger variations in the cycle-averaged values across consecutive revolutions. This variance was particularly pronounced for the high-density mesh, which was prone to abrupt, significant fluctuations in thrust that made it difficult to determine a logical termination point for the simulation. As mentioned previously, modelling vortex decay in 2D simulations is challenging and frequently relies on artificial numerical diffusion to compensate. Consequently, in the absence of 3D instability modelling, reducing numerical diffusion in 2D may produce unrealistically persistent coherent vortical structures, leading to extended blade-vortex interactions and reduced periodic repeatability. Since the optimisation objective depends on cycle-averaged performance, the 800k mesh was selected as a compromise between resolution, periodic repeatability, and computational cost. As a final check, the initial design space for the four-bladed cyclorotor aerofoil optimisation was also simulated with the 4,000k mesh and the cross-correlation coefficient between $FM$ values using the 800k and 4,000k mesh was found to be 0.96, indicating strong agreement in design trends. 
	
	\section{Vortex characteristics during dynamic stall}
	
	To support the discussion of LEV behaviour across varying blade counts and aerofoil geometries, coherent vortices were identified over a cycle using the $\Gamma_2$ criterion with a threshold of $2/\pi$ \citep{Graftieaux2001CombiningFlows,DeGregorio2020AnData}. The centroid of each identified region defines the instantaneous vortex core position, and the local vorticity was integrated over this bound area to quantify the vortex circulation.
	
	\begin{figure}[h]
		\centering
		\includegraphics{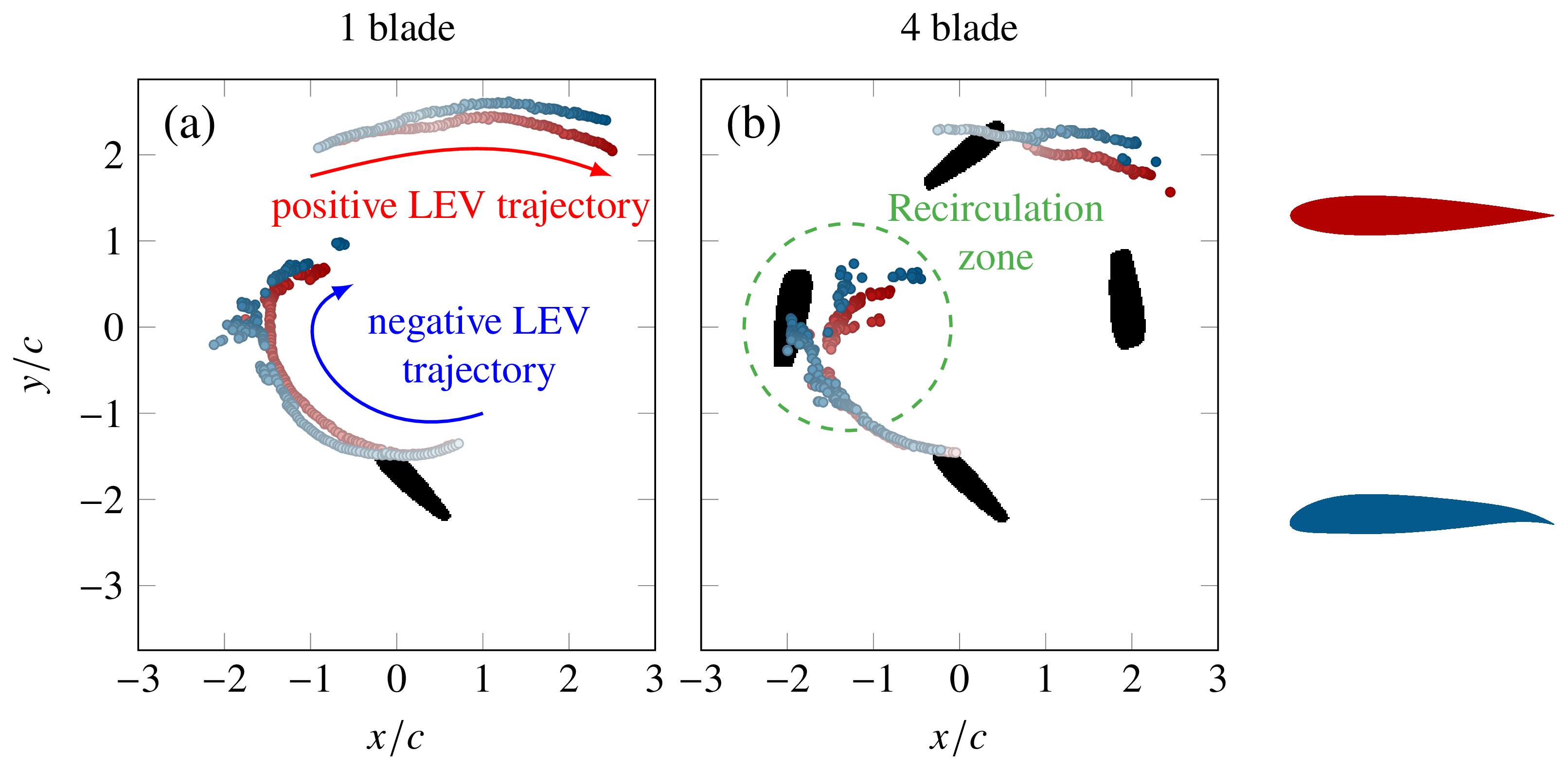}
		\caption{Vortex positions found via $\Gamma_2$ criterion over single revolution for 1-blade (left) and 4-blades (right) with both the baseline and optimised aerofoils. Marker-colour darkens with time.}
		\label{fig:vortex_trajectories}
		\phantomsubcaption\label{fig:vortex_trajectories_A}
		\phantomsubcaption\label{fig:vortex_trajectories_B}
	\end{figure}
	
	A comparison of the vortex trajectories reveals that the positive and negative LEVs follow similar trajectories irrespective of blade count or aerofoil profile (Fig. \ref{fig:vortex_trajectories}). Negative LEVs form within the lower half of the rotor trajectory, following the blade path and finally detaching within the recirculation zone. Following detachment, these vortices lose coherence within this zone.  This results in a highly scattered distribution of tracked positions as the initially coherent LEV disintegrates into multiple secondary vortices. Conversely, positive LEVs form within the upper half of the rotor circle and advect rightwards following detachment. All vortex convection speeds are approximately equal to the rotational velocity of the blades upon which they formed. The one exception to this are the negative LEVs after detachment which remain largely stationary within the recirculation. 
	
	In the 1-blade configuration, negative LEVs develop earlier in the cycle than in the 4-blade case. This observation supports the interpretation that the induced throughflow inherent to the 4-blade configuration suppresses the effective angle of attack, thereby delaying the threshold angle-of-attack required for vortex formation. Throughflow effects are equally visible along the positive LEV trajectories, which exhibit a more pronounced downward deflection as they are advected by the rotor induced throughflow.
	
	For the 1-blade configuration, the optimised aerofoil generates a negative LEV trajectory similar to that of the baseline case, albeit marginally shifted radially outward, indicating that the vortex remains more closely bound to the blade suction surface (Fig. \ref{fig:vortex_trajectories_A}). Similarly, the corresponding positive LEVs follow a path shifted further radially outward. These differences in vortex trajectories between the baseline and optimised profiles persist in the 4-blade configuration (Fig. \ref{fig:vortex_trajectories_B}).

	\begin{figure}[h]
		\centering
		\includegraphics{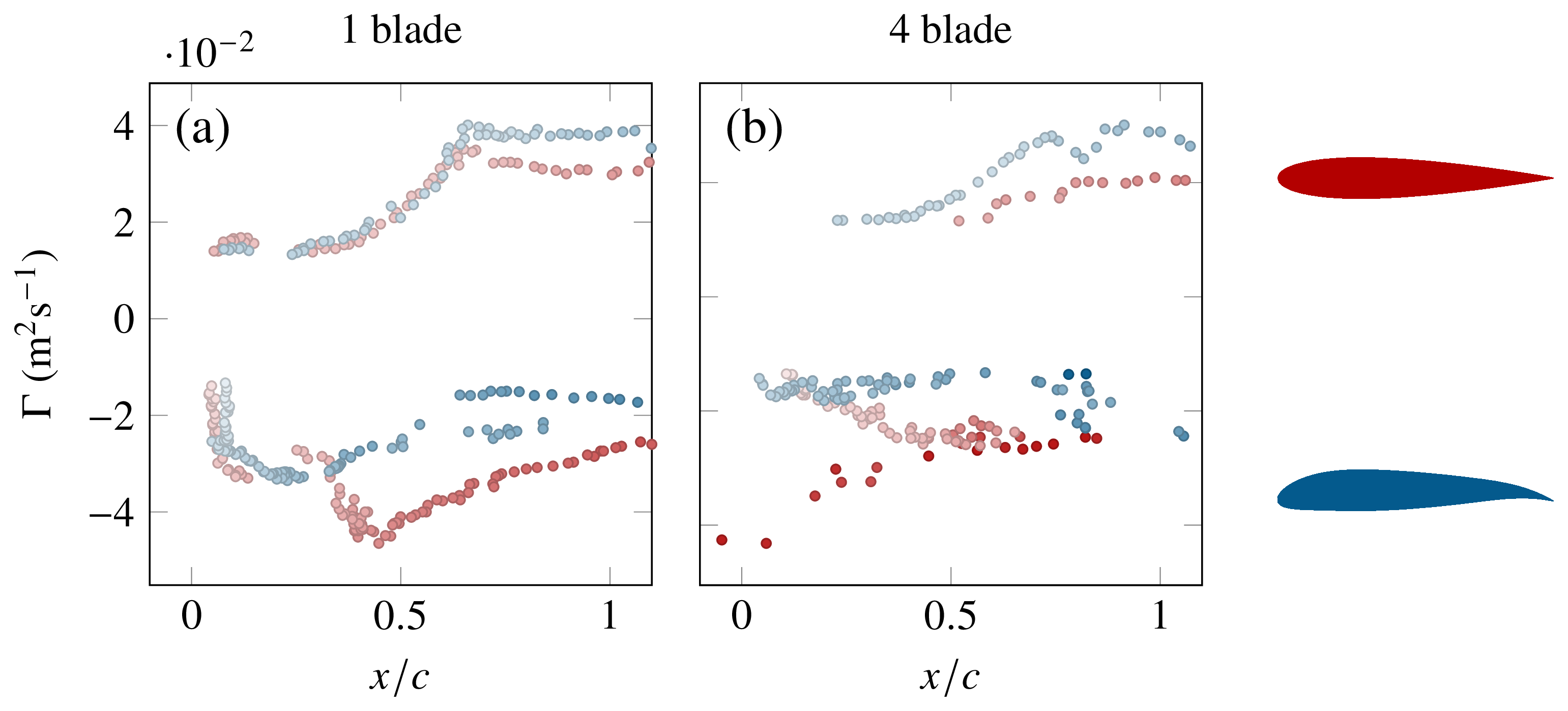}
		\caption{Circulation within vortices found via $\Gamma_2$ criterion in a blade-fixed reference frame over single revolution for (a) 1-blade and (b) 4-blades with both the baseline and optimised aerofoils.  Marker-colour darkens with time.}
		\label{fig:vortex_circulation}
		\phantomsubcaption\label{fig:vortex_circulation_A}
		\phantomsubcaption\label{fig:vortex_circulation_B}
	\end{figure}

	The LEV dynamics can be further quantified by evaluating the circulation as a function of chordwise position within a blade-fixed reference frame (Fig. \ref{fig:vortex_circulation}). In the 1-blade configuration, both LEVs initiate near the leading edge (LE) and convect toward the trailing edge (TE), growing to a peak circulation value before undergoing breakdown and decay (Fig. \ref{fig:vortex_circulation_A}). With the baseline aerofoil, the negative LEV grows more rapidly and achieves a higher peak circulation than its positive counterpart.
	
	The optimised aerofoil reduces the negative LEV's peak circulation significantly, confirming that improved boundary layer attachment limits the magnitude of vorticity shed into the wake. The optimised aerofoil's positive LEV behaves similarly to the baseline case but exhibits a slightly higher peak circulation. This occurs because the leading-edge droop of the optimised profile is aerodynamically inverted relative to the oncoming flow during this portion of the cycle, effectively increasing the local angle of attack and promoting flow separation.
	
	The 4-blade circulation data is inherently more complex due to multi-blade wake interactions, making it challenging to isolate the specific vortices associated with a single reference blade (Fig. \ref{fig:vortex_circulation_B}). Nonetheless, the data mirrors the 1-blade trends at lower peak magnitudes. This reduction in vortex intensity supports the earlier conclusion that higher blade counts strongly moderate the effective angle-of-attack and suppresses dynamic-stall severity.

	\bibliographystyle{apalike}
	\bibliography{references}
	
\end{document}